\documentclass[superscriptaddress,floatfix,showpacs,preprint,prb] {revtex4}

\usepackage[normalem]{ulem}

\usepackage{amsmath,amssymb,epsfig,color,bm,graphicx,alltt}
\usepackage{amsfonts}
\usepackage{latexsym}
\usepackage{wrapfig}
\usepackage{hyperref}

\usepackage{grffile}

\newcommand{\I}{{\rm i}}

\renewcommand{\ge}{\geqslant}
\newcommand{\kk}{\mathbf{k}}

\newcommand{\mf}{{m_{\rm FM}}}
\newcommand{\mfmf}{{m^2_{\rm FM}}}

\newcommand{\muc}{{\mu_{\rm PS}}}

\begin{document}

\title{Inverse magnetocaloric effect and phase separation induced by~giant van Hove singularity in itinerant ferromagnetic metal}

\author{P.\ A.\ Igoshev\thanks{E-mail: igoshev\_pa@imp.uran.ru}}

\affiliation{M.~N.~Mikheev Institute of Metal Physics, 620041, S. Kovalevskoj Str. 19, Ekaterinburg, Russia}

\affiliation{Amirkhanov Institute of Physics of DFRC of RAS, Makhachkala 367003, Russia}

\author{I.\ A.\ Nekrasov\thanks{E-mail: nekrasov@iep.uran.ru}}

\affiliation{Institute of Electrophysics UB RAS, 620016, Amundsena Str. 106, Ekaterinburg, Russia}


\begin{abstract}
A thermodynamic theory based on Landau grand potential expansion   
for ferromagnetic-paramagnetic phase transitions is developed for an electronic phase-separated state. 
It is rigorously shown that ferromagnetic phase involved in the phase-separated state exhibits negative magnetic susceptibility in the vicinity of~tricritical point. 
Thus, an entropy of the magnetically ordered phase 
may increase when the magnetic field is applied, which implies positive sign of the total magnetic entropy change $\Delta S$ within magnetocaloric effect~(MCE). 
The electronic phase separation and MCE are considered within the Hubbard model for~face-centered cubic lattice with giant van Hove singularity of electron density of states at the band bottom. 
Within the Hartree-Fock approximation it is shown that such model of itinerant magnet exhibits the~first-order ferromagnet-paramagnet phase transition~(FOPT)  with electronic phase separation and inverse magnetocaloric effect deep inside the phase-separated region. 
Temperature dependence of $\Delta S$ for the mean-field solution of the non-degenerate Hubbard model is analyzed in detail for different band filling values. 
The possibility to control $\Delta S$ sign by changing both temperature and band filling of magnetocaloric materials is demonstrated. This is important to interpret a lot of experimental data, possible technological applications, and further theoretical developments.
\end{abstract}
\maketitle

\section{Introduction}\label{sec:Introduction}
The magnetocaloric effect~(MCE), i.e.,~a change of a~sample temperature~$T$~(entropy $S$) when the external magnetic field is~applied under the condition of fixed entropy~(temperature), is observed in a lot of magnetically ordered systems. The strength of MCE typically has maximal value close to critical temperature of magnetic phase transition. 
Over the past 40 years, a huge amount of experimental data on MCE has been accumulated, see reviews~\onlinecite{Franco2012, Li2016,2018:Franco}. 
The current state of affairs in the field of materials used for magnetic cooling is described in Ref.~\onlinecite{Tishin14}, whereas medical applications are reviewed in~Ref.~\onlinecite{Tishin2016}. 
Details of standard theoretical approaches to study of MCE  are described in the review~\onlinecite{Oliveria10}.

The temperature profile of isothermal entropy change $\Delta S(T)$ and its maximum absolute value $\Delta S_{\rm max}$ substantially depend on the order of magnetic phase transition as well as on~the~nature of the electronic states forming magnetic order. 
Most important quantity for applications characterizing the magnetic cooling efficiency is \textit{relative cooling power} proportional to $\Delta S_{\rm max}$, and its peak width. 
Thus, an ability to calculate $\Delta S(T)$ profile for~real materials based on characteristics of their electronic structure is of a great importance for wide practical applications.  
In addition, not only a large value $\Delta S$ at peak temperature, but also the table-like temperature dependence of $\Delta S$ can be used to construct an ideal Ericsson cycle of magnetic refrigeration. Recently the compounds exhibiting such behavior attracted a lot of attention~\cite{2014:Ericsson:Gd-Co-Al,2015:Ericsson:Gd56Ni15Al27Zr2,2016:Ericsson:Gd50Co45Fe5,2017:Ericsson:Gd-Ni-Al,2018:Ericsson:GdMnSi}.

Perhaps the most known material being examined for practical use is rare-earth system Gd$_5$Si$_2$Ge$_2$ exhibiting so-called giant MCE ($\Delta S_{\rm max}\sim 19$~J/kg/K under the magnetic field 5~T) at room temperature~\cite{Pecharsky97}.  The magnetic properties of this system are mainly driven by localized magnetic moments of Gd $f$ shell. For this compound, the position of~$\Delta S(T)$ maximum is close to the Curie temperature $T_{\rm C}\sim273$~K of the first-order ferromagnetic~(FM)--paramagnetic~(PM) phase transition~(FOPT).
This FOPT is accompanied by multiphase inhomogeneous crystal structure and chemical composition in the Gd$_5$Si$_2$Ge$_2$ doped by Ga~\cite{2008:Kumar:Gd5SiGe2_doped_Ga}.
Despite outstanding MCE characteristics Gd$_5$Si$_2$Ge$_2$ is rather expensive for wide practical applications.

Rare-earth compounds are not the only materials with remarkable MCE properties. There are some examples of transition-metal compounds, which also exhibit large and unexplained MCE potential values. The MnAs, MnFeP$_{0.45}$As$_{0.55}$, and La(Fe$_x$S$_{1-x}$)$_{13}$ systems demonstrate $\Delta S_{\rm max}$ value close to that of the  Gd$_5$Si$_2$Ge$_2$ system. 
Note that, first-order FM-PM phase transition in the following compounds occurs at temperature close to room temperature: 
318~K in MnAs~\cite{MnAs_2002_FM}, about 300~K in MnFeP$_{0.45}$As$_{0.55}$~\cite{MnFePAs_2002_FM}.
First-order FM-PM phase transition occurs at a temperature about 195~K in~La(Fe$_{0.88}$Si$_{0.12}$)$_{13}$ and its origin is likely the peak of electron density of states in the vicinity of Fermi level in paramagnetic phase~\cite{Fujita2003}.  
Interestingly,  the order of FM--PM magnetic phase transition in the system La(Fe$_x$Si$_{1-x}$)$_{13}$ depends on the concentration of Si: at low concentrations there is a first-order phase transition, while at larger concentrations second-order phase transition (SOPT) occurs~\cite{Franco2017}.  
Magnetism in the systems MnFeP$_{0.45}$As$_{0.55}$ and La(Fe$_x$Si$_{1-x}$)$_{13}$ has pronounced itinerant character~\cite{2003:Yamada}. 

All above-mentioned compounds demonstrate so-called \textit{direct} MCE: an increase of the magnetic field within an isothermal process results in the entropy decrease ($\Delta S < 0$), and the temperature increase within the adiabatic process ($\Delta T>0 $). 
However, there is another class of compounds exhibiting opposite response to an applied magnetic field ($\Delta T < 0$, $\Delta S > 0$), which is~named \textit{inverse}~MCE. 

Typical classes of systems with inverse MCE~($\Delta T < 0$, $\Delta S > 0$) are Heusler alloys (\textit{e.g.},~NiMnIn~\cite{inverse_direct_PRB}, NiMnSn~\cite{NiMnSn_2005}, Ni$_{2}$Mn$_{1-x}$Cu$_{x}$Ga~\cite{strange_ferri_anom} with martensitic antiferromagnetic~(AFM)--FM phase transition), some manganites (\textit{e.g.},~Pr$_{0.5}$Sr$_{0.5}$MnO$_3$~\cite{Hsini2018},  La$_{1-x}$Ca$_x$MnO$_3$~\cite{Restrepo-Parra2014}),  intermetallic compounds (\textit{e.g.}~PrNi$_5$~\cite{PrNi5_1998}, ErRu$_2$Si$_2$~\cite{ErRu2Si2_Samanta2007}, CoMnSi$_{1-x}$Ge$_x$~\cite{CoMnSiGe_Sandeman2006}), for more details see the reviews~\cite{Gschneider05, Franco2012, Li2016, Oliveria10}. 
In the Heusler alloys, the FM--PM magnetic phase transitions typically give direct MCE while inverse MCE occurs under FM-AFM transition.

Besides observed above ordinary ferromagnetic systems, 
there are also systems where the ferromagnetic order coexists with some other ordered phases in terms of~phase separation~(PS): 
MnFeP$_{0.8}$Ge$_{0.2}$, Mn$_{0.99}$Cu$_{0.01}$As~\cite{CARON20093559}~(FM$+$PM phase separation), 
La$_{0.27}$Nd$_{0.40}$Ca$_{0.33}$MnO$_3$\cite{PS_material_anom} (FM$+$charge-ordered phase separation), Gd$_{5}$Ge$_{2.3}$Si$_{1.7}$\cite{PS_Mce}~(FM$+$AFM phase separation). 
The compound (alloy) FeRh exhibits large value of $\Delta S_{\rm max}>0$ (inverse MCE) under corresponding FM-AFM FOPT upon temperature decrease  with critical temperature 270~K~\cite{FeRh,Vaulin}. This phase transition goes through FM$+$AFM phase separation region. 

There are also some materials that exhibit inverse MCE: MnRhAs~\cite{ferri_anom}, Co and Mn Co-doped Ni$_2$MnGa~\cite{strange_ferri_anom}, Er$_2$Fe$_{17}$~\cite{KHEDR2019436,PhysRevB.86.184411}, DyAl$_2$\cite{PhysRevB.61.447}, Pr$_{0.46}$Sr$_{0.54}$MnO$_3$\cite{anomal_fail1st} in the vicinity of FOPT temperature.  Typically, there are two different temperatures for inverse and direct MCE. Moreover, an inverse MCE provides an additional peak in $\Delta S$ temperature profile (see, e.g.,~Eu$_{0.55}$Sr$_{0.45}$MnO$_3$~\cite{PS_material_anom},  Ni$_{50}$Mn$_{33.13}$In$_{13.90}$~\cite{inverse_direct_PRB},ErGa$_2$, HoGa$_2$~\cite{inverse_direct_anisotropy},  HoFeSi~\cite{successive_anom_direct_1st}).

Generally, the magnetic materials can be divided onto two classes: localized magnetic moment and itinerant magnets. 
In the first class the corresponding electron states are well localized, typically derived from $f$-electron states and can be described by electron spin only. 
In this case, magnetic degrees of freedom for an electron state is a spin projection only and magnetic interaction is some kind of exchange interaction between spins at different lattice sites. 
In the second class there is an additional degree of freedom, a position of electron state in the Brillouin zone, indexed by~a~quasimomentum~$\mathbf{k}$. 
A factor favoring to itinerant magnetism is an accumulation of extremely large number of electron states in the vicinity of Fermi level due to van Hove singularity. 
Van Hove singularity (vHS) is possibly an origin of ferromagnetic ordering in $\alpha$ iron~\cite{2015:Igoshev:Fe}, ZrZn$_2$~\cite{2020:Skornyakov:ZrZn2}, nickel~\cite{2017:Hausoel:Ni}, Ni$_3$Al~\cite{2011:Hamid:Ni3Al},  LuCo$_3$~\cite{2020:LuCo3:Neznahin,2022:LuCo3:Radzivonchik}.  
In the absence of vHS near~the~Fermi level,  ferromagnetism formation by an itinerant scenario is seemingly impossible. 
While in two-dimensional case logarithmic vHS is protected by van Hove theorem, in three-dimensional case there should be additional properties (higher-order vHS, van Hove singularity lines) of vHS to provide a divergence or valuable feature of density of electron states to stabilize ferromagnetism~\cite{1993:VonsovskiiI,1993:VonsovskiiII,2019:Igoshev_PMM,2019:Igoshev_JETP_letters}. Higher-order vHS are now attracting high attention due to relevance to graphene properties~\cite{2019:LYuan,2022:Guerci:vHS_in_graphene}.  

The FOPT typically is accompanied by some sample heterogeneity (phase separation).
In localized electron magnets, phase separation is manifested in charge density uniform sample which contains regions with different composition and crystal structures. To explain it one needs to take into consideration some additional degrees of freedom such as boundary effects, crystal structure transition, magnetoelastic coupling, and so on. 

For the itinerant systems there appears additional scenario of phase separation---pure \textit{electronic} phase separation~[52]. The electronic phase separation manifests itself by regions with different conducting electron charge density while lattice structure may be the same for the whole sample. However, such conducting electron charge density redistribution may be followed by crystal structure transition. 

Here in our paper we consider only the electronic phase separation scenario, which is physically quite different from phase separation in localized magnetic moment systems but requires its own theoretical development. 

Previously some theoretical studies of MCE in materials and corresponding models with FOPT were performed. 
A Monte-Carlo-based solution of the Potts-Blume-Emery-Griffiths model was used to investigate a series of Ni$_{45}$Co$_{5}$Mn$_{37}$In$_{13}$~\cite{Comtesse2014}
and Ni-Co-Mn-(Sn,Al) alloys~\cite{Sokolovskiy2018}, which have two phase transitions with direct and inverse MCE. 
Within the Potts-Blume-Emery-Griffiths model, the inhomogeneity arising during the FOPT was taken into account by introducing an additional term to the Hamiltonian responsible for a disorder, which~simulates the mixing of martensitic and austenitic phases. Investigation of MCE for the~Hubbard model within Hartree-Fock approximation for different lattices under the assumption that phase transition of the second order  was performed in~Refs.~\onlinecite{Oliveria10,2017:Igoshev}.  
Theoretical study of MCE for FM-AFM FOPT within the same model and approximation taking into account both ferrimagnetic phases and the possibility of phase-separated state formation was performed for Bethe~\cite{2021:Ivchenko} and square~\cite{2023:Igoshev_MCE_PS_square_lattice} lattices. 

Another direction of theoretical studies is based on phenomenological Landau theory. The latter was designed to describe the FM-PM SOPT using a simple free energy functional~\cite{Landau}. 
However, this theory opens up wide possibilities for considering more general cases, \textit{e.g.},~the~case of FOPT. 
The Landau theory equally well can be applied to both itinerant and localized electron systems.  
For the itinerant systems a~direct derivation of Landau series coefficients within the Stoner approximation can be done straightforwardly using bare electronic density of states~\cite{Moriya}. 
Peculiarities of electronic density of states (DOS) were considered as an origin of FOPT within Landau based theory of itinerant metamagnetism, see,~e.~g.,~Refs.~\onlinecite{1992:Duc,1981:Shimizu}. 
However, another consequence of FOPT in itinerant systems may be an electronic phase separation~\cite{2021:Kagan} which serves as~a~focus of our study in the context of MCE. 
Some examples of the Landau theory application for explanation of FOPT and MCE in itinerant systems Co(S,Se)$_2$, Lu(Co,Al)$_2$, and Lu(Co,Ga)$_2$ can be found in~Ref.~\onlinecite{2003:Yamada}.  
Necessary for FOPT sign changes of Landau series coefficients may be provided due to~the~presence of peculiarities of density of states in the vicinity of the Fermi level~\cite{Fujita2003} or due to~an~introduction of additional interactions (see studies for, e.~g., AB$_2$~compounds~\cite{2003:Yamada},  MnAs~(\onlinecite{2004:Oliveira,2013:Oliveira,Oliveria10}), MnFeP$_{0.45}$As$_{0.55}$~(\onlinecite{2005:Oliveira}), YCo$_2$~(\onlinecite{1975:Bloch,1992:Duc})). However, the electronic phase separation accompanying FOPT in itinerant systems stays beyond such approach and requires further extension given in this paper. 

Thus, there is a problem of theoretical modeling and explanation of MCE in itinerant systems exhibiting FM-PM FOPT . 
Another complexity comes from necessary presence of phase-separated state in itinerant systems near FOPT region, and pure localized electron model (without itinerant component) cannot consistently treat appearing of inhomogeneities~\cite{2009:Balli_wrong_Maxwell,2009:Pecharsky_Phase-separated_state,XU20153149,2013:Oliveira,otherGMCEstudy,1storder,1325142}. 
At the moment there is a lack of theoretical studies of MCE for variety of experimental data in systems with FOPT and PS. 
An additional issue is that conventional Maxwell relations for experimental data treatment in a case of FOPT should be applied with a great caution.
Here, one should take into account that phase volumes within the PS state depend on temperature and magnetic field~\cite{PS_material_anom}.  Neglecting this dependency leads to spurious results~\cite{Pecharsky97,2009:Balli_wrong_Maxwell}, which were discussed in Refs.~\onlinecite{Comtesse2014,wrong_Maxwell}. 

The plan of the presentation is the following. 
An extension of the Landau theory for an~account of~PS state during FM-PM FOPT is presented in Section~\ref{sec:Landau_theory}. 
Section~\ref{sec:calculation_example} contains the~MCE~study within~the~mean-field solution of the non-degenerate Hubbard model for~the~face-centered cubic (fcc) lattice with the parameter set corresponding to the position of Fermi level in the vicinity of giant van Hove singularity of the electronic density of states. 
The paper is ended by concluding remarks in Sec.~\ref{sec:conclusions}.

\section{FM-PM FOPT thermodynamics: the extension of Landau theory}\label{sec:Landau_theory}

At fixed temperature in energy units $T$, band filling $n$, magnetic field in energy units $h = \mu_{\rm B}H$ with $\mu_{\rm B}$ being the Bohr magneton, $H$ being a magnetic field, conventional free energy Landau functional $F(T,n,h|m)$ is a functional of the magnetization per site (in~units~of~$\mu_{\rm B}$) $m$ playing a role of an~order parameter. However functional $F$ does not allow us to describe electronic phase separation (PS) which can accompany any FOPT in itinerant magnets~\cite{2021:Kagan,2010:Igoshev,2015:Igoshev_JPCM}. 
From a general point of view, in itinerant magnet order parameter jump at FOPT point should result in~corresponding band filling jump, which in turn results in PS formation.  
To take this into account, a~chemical potential $\mu$ should be introduced instead of $n$ as an~argument of the Landau functional and a~grand potential $\Omega$ instead of free energy $F$~\cite{Landau,2023:Igoshev_PMM}. 
In this section we present Landau theory based thermodynamical approach for FM-PM phase transition in an itinerant system  including both FOPT (with PS) and SOPT cases on equal footing. 

Let us write Landau expansion for grand potential $\Omega $ with respect to the order parameter~$m$
\begin{equation}\label{eq:Landau_exp}
	\Omega(T, \mu, h | m) = A^\Omega_0(T, \mu) + A^\Omega_2(T, \mu)m^2 + A^\Omega_4(T, \mu)m^4 + A^\Omega_6(T, \mu)m^6 - h m,
\end{equation}
where $A^\Omega_6 >0$ provides a stable solution. 
The coefficients $A^\Omega_\alpha$ up to sixth order are held to supply not only SOPT but also the~possibility of~FOPT. 

External arguments (\textit{i.e.}, set up by environment such as $T$, $\mu$, and $h$) are written before the vertical line `$|$'; the internal argument, magnetization $m$, which can be found by minimization of~$\Omega$ under given external parameters, is written after `$|$'. 
Note that $m$ value specifies particular phase for the system. 
We consider phase transitions between phases differing by a~value of $m$ only, \textit{e.~g.}~paramagnetic-ferromagnetic or~antiferromagnetic-ferrimagnetic phase transitions.  

The \textit{equation of state} defining magnetization $m= m_\Phi(T,\mu,h)$~($\Phi$ denotes particular phase) for~given $T, \mu$ and $h$ is determined by extremum condition of Eq.~\eqref{eq:Landau_exp} $\partial \Omega/\partial m = 0$, and has a~form
\begin{equation}\label{eq:implicit_eq_state}
	f(T, \mu | m) = h,
\end{equation}
where
\begin{equation}
	 f(T, \mu | m) =  2 \left(A^\Omega_2(T, \mu) m + 2 A^\Omega_4(T, \mu) m^3 + 3 A^\Omega_6(T, \mu) m^5\right)
\end{equation}
and $\partial f/\partial m > 0$ (the~criterion of $\Omega$ minimum at~$m$~point). 

We are interested in the calculation of zero-field magnetic susceptibility $\chi^{\rm env}_\Phi =  \left(dm_\Phi/dh\right)_{h = 0}$
\begin{equation}\label{eq:chi_general}
    \chi^{\rm env}_\Phi = \bar{\chi}_\Phi + \eta_{\rm env}(h = 0)\left(\frac{\partial m_\Phi}{\partial\mu}\right)_{h = 0}, 
\end{equation}
where $\mu$-fixed (as denoted by bar) zero-field magnetic susceptibility $\bar{\chi}_\Phi = \left({\partial m_\Phi}/{\partial h}\right)_{h = 0}$, the derivative 
\begin{equation}\label{eq:eta_def}
\eta_{\rm env} = d\mu/dh
\end{equation}
originates from the dependence $\mu = \mu(h)$,  which in turn must be specified to enforce the constraints imposed by the physical environment (``env''). 
Using Eq.~(\ref{eq:implicit_eq_state}) we get  
$\bar\chi = \left({\partial f}/{\partial m}\right)^{-1}_{h = 0, m = m_\Phi}$, i.~e., 
\begin{equation}\label{eq:dmdh}
	\bar\chi_\Phi  = \frac12(A^\Omega_2 + 6 A^\Omega_4 m^2_\Phi + 15 A^\Omega_6 m^4_\Phi)^{-1},
\end{equation}
where $m = m_\Phi$ satisfies Eq.~(\ref{eq:implicit_eq_state}) at $h = 0$. 
At $h = 0$, the~Eq.~(\ref{eq:implicit_eq_state}) may have up to~two solutions~$m\ge0$. 
While in the case $A^\Omega_2<0$ unique minimum exists at arbitrary values of $A^\Omega_4$, 
at $A^\Omega_2>0$, $m = m_{\rm PM} = 0$ $\Omega$ minimum always exists (corresponding to~PM~state) and  
in the case $-\sqrt{3 A^\Omega_2A^\Omega_6} < A^\Omega_4 < 0$, additional finite-$m$ minimum of $\Omega$ exists, $m = m_{\rm FM}$,   
\begin{equation}\label{eq:explicit_eq_state}
	m^2_{\rm FM} = \frac{-A^\Omega_4 + \sqrt{\left(A^\Omega_4\right)^2 - 3 A^\Omega_2A^\Omega_6}}{3A^\Omega_6}.
\end{equation}
corresponding to FM state. 

For magnetically ordered phase, we calculate $\bar\chi$, taking into account Eqs.~(\ref{eq:implicit_eq_state}) and (\ref{eq:dmdh}),
\begin{equation}\label{eq:chi_m_FM}
	\bar\chi_{\rm FM} = \frac1{8m^2_{\rm FM}(A^\Omega_4 + 3A^\Omega_6m^2_{\rm FM})}.
\end{equation}
For PM solution, from~Eq.~(\ref{eq:dmdh}) we have 
\begin{equation}\label{eq:chi_m_PM}
	\bar\chi_{\rm PM} = \frac1{2A^\Omega_2}.
\end{equation}

Further we consider two different cases: (i) single-phase~(homogeneous)  and (ii) phase-separated state. 
\subsection{Magnetic susceptibility at fixed band filling}
For a single-phase case (i)~the~band filling function $\mathfrak{n}(\mu, T | m) =  -\partial \Omega/\partial \mu$ is calculated as  
\begin{equation}\label{eq:n_exp}
\mathfrak{n}(T, \mu | m) = -(\dot{A}^\Omega_0 + \dot{A}^\Omega_2m^2 + \dot{A}^\Omega_4m^4 + \dot{A}^\Omega_6m^6), 
\end{equation} 
with the notation $\dot{A}^\Omega_\alpha = \partial{A^\Omega_\alpha}/\partial\mu$.   
Therefore for the single-phase state there comes additional $h$-independent condition to determine $\mu$ using given band filling value~$n$
\begin{equation}\label{eq:n=const}
	\mathfrak{n}(T, \mu | m) = n
\end{equation}
as a function of $T$ and $m$.  A magnetic field $h$ affects the system only through $m$~and~$\mu$.

The fixed band filling magnetic susceptibility (as denoted by upper index ``$n$'') reads
\begin{equation}
\label{eq:chi_def_n}
\chi^n_\Phi \equiv \left(\frac{dm_\Phi}{dh}\right)_{n, h = 0}
\end{equation}
is obtained directly from Eq.~(\ref{eq:chi_general}), 
where $\mu$ changes consistently with $m$ to satisfy Eq.~(\ref{eq:n=const}), so that $\eta_n = \left({d\mu}/{dh}\right)_{n,h=0}$, which results in
\begin{equation}\label{eq:eta_def_n}
	\eta_n = -\left(\frac{{\partial \mathfrak{n}}/{\partial m}}{{\partial \mathfrak{n}}/{\partial \mu}}\right)_{h = 0,m = m_\Phi}\chi^n_\Phi.
\end{equation}

Therefore we get 
\begin{equation}\label{eq:chi_n}
	\chi^n_\Phi = \left(\frac{\partial f}{\partial m} - \frac{\partial f}{\partial \mu}\frac{{\partial \mathfrak{n}}/{\partial m}}{{\partial \mathfrak{n}}/{\partial \mu}}\right)^{-1}_{h=0,m = m_\Phi}
\end{equation}
which is always positive for~any~single-phase itinerant electron system. 

\subsection{Magnetic susceptibility of particular phases for phase-separated state}

Now we consider the case (ii)~for phase-separated state caused by FOPT with PM and FM phases (instead of~these phases there can be any other pair of phases differing from each other only by the presence of non-zero component of total magnetization). 
\begin{figure}
	\includegraphics[angle=-90,clip=true,width=0.79\textwidth]{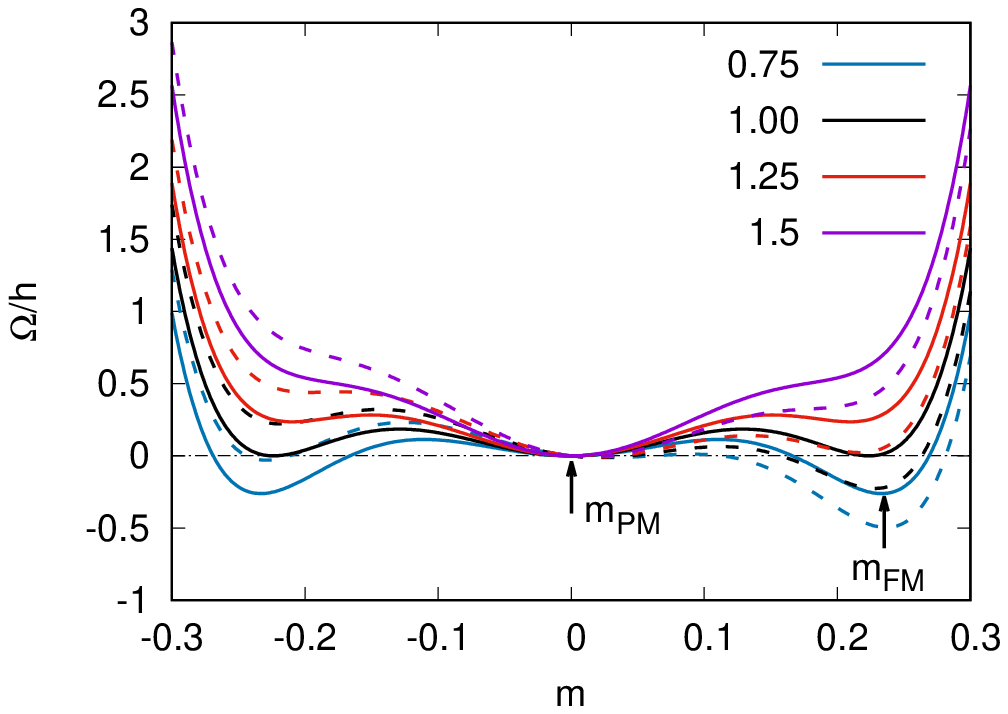}
\caption{
\label{fig:m_vs_Omega}
    Schematic plot of $\Omega(m)$ in units of $h$, see Eq.~\eqref{eq:Landau_exp}, for $A^\Omega_4 = -0.1h$, $A^\Omega_6 = h$, and different values of $A^\Omega_2$ in units of $A^\Omega_{2,\rm FOPT}$~(shown in legend). Arrows show the positions of  minimum of $\Omega(m)$, corresponding to minimum position $m = m_{\rm FM}(m_{\rm PM})$ for~FM~(PM) phase. 
    Solid  (dashed) lines show the result at zero (finite, equal to $h$) magnetic field.}
\end{figure}
Main difference of FOPT in itinerant systems from the FOPT in localized systems is that in~the~former case, a~jump of the order parameter $m$ (magnetization) at~the~phase transition point leads to a band filling jump (electronic PS). 
For the phase-separated state total band filling $n$ is a~weighted sum of both phase band fillings $n_{\rm FM}$, $n_{\rm PM}$, see Eq.~\eqref{eq:n_exp}, satisfying Eq.~(\ref{eq:implicit_eq_state}) at the thermodynamic equilibrium.  
For given $T$ and $h$ at FOPT line the criterion of PS appearance is a falling of $n$ into the band filling jump interval between $n_{\rm FM}$, $n_{\rm PM}$, see details in~Refs.~\onlinecite{2010:Igoshev,2021:Ivchenko,2023:Igoshev_PMM,2015:Igoshev_JPCM}. 
Therefore all extensive quantities describing PS state acquire additional linear dependence on band filling $n$.
Thus, for $\mu$ fixed at FOPT line band filling $n$ plays a role of additional degree of freedom.

At FOPT line, grand potential $\Omega(\mu, T, h | m_\Phi)$ for different phases $\Phi=\text{FM},\text{PM}$ involved in~PS should be equal,  where $m_{\rm FM}$ and $m_{\rm PM}$ are the roots of Eq.~(\ref{eq:implicit_eq_state}). 
The definition of phase separation boundary line $\mu = \muc(T,h)$ reads
\begin{equation}\label{eq:PS_condition}
	\Omega(T, \muc(T,h), h | m_{\rm FM}) = \Omega(T, \muc(T,h), h | m_{\rm PM}).
\end{equation}

In~the~case of zero magnetic field $h = 0$, Eq.~(\ref{eq:PS_condition}) yields
\begin{equation}\label{eq:explicit_eq_state_simple}
	\mfmf = -\frac{2A^\Omega_2}{A^\Omega_4}.
\end{equation}
Using the Eq.~(\ref{eq:explicit_eq_state}) one can get zero-field FOPT line equation in terms of $T$ and $\mu$ variables
\begin{equation}\label{eq:1st_order_A_condition}
	 A^\Omega_2(T,\mu) = A^\Omega_{2,\rm FOPT}(T,\mu),
\end{equation}
where 
\begin{equation}\label{eq:A2_FOPT}
    A^\Omega_{2,\rm FOPT}(T,\mu) = \frac{\left(A^\Omega_4(T,\mu)\right)^2}{4A^\Omega_6(T,\mu)}.
\end{equation}
Figure~\ref{fig:m_vs_Omega} shows a schematic example of $m$ dependence of $\Omega(T,\mu, h|m)$ at zero and finite magnetic field, see~Eq.~\eqref{eq:Landau_exp}, when the~coefficient $A^\Omega_2$ goes over $A^\Omega_{2,\rm FOPT}$ value. One can see that at $h = 0$ global minumum position of $m$ dependence of $\Omega(T,\mu, h|m)$ changes from $m = m_{\rm FM}$ to $m = m_{\rm PM}$ as $A^\Omega_2(T,\mu)$ becomes larger than $A^\Omega_{2,\rm FOPT}(T,\mu)$. At $A^\Omega_2(T,\mu) > (4/3)A^\Omega_{2,\rm FOPT}(T,\mu)$ local FM minimum disappears.  

Let $A^\Omega_4 \rightarrow 0$ at FOPT line whereas $A^\Omega_6$ remains finite, so from  Eqs.~\eqref{eq:explicit_eq_state_simple} and \eqref{eq:A2_FOPT}  we get $\mfmf,A^\Omega_2 \rightarrow 0$ and we arrive at \textit{tricritical} point ($T = T_{\rm TCP}$, $\mu = \mu^\ast$) for which $A^\Omega_2(T, \mu) = A^\Omega_4(T, \mu) = 0$. The continuation of FOPT line (Eq.~(\ref{eq:1st_order_A_condition})) corresponding to~equation~$A^\Omega_2(T, \mu) = 0$ in $A^\Omega_4(T, \mu) > 0$ region yields SOPT line. Thereby, at~tricritical point phase transition changes its order.

Using the~Eqs.~(\ref{eq:explicit_eq_state_simple}) and~(\ref{eq:1st_order_A_condition}), we simplify the expression (\ref{eq:chi_m_FM}) for $\mu$-fixed FM phase magnetic susceptibility
\begin{equation}\label{eq:bar_chi_FM}
	\bar\chi_{\rm FM} = \frac{1}{8A^\Omega_2} > 0,
\end{equation}
which is essentially positive in case of FOPT. 
Obtained results are 
applicable for both itinerant and localized (with replacements $\mu\rightarrow n$ and $\Omega\rightarrow F$) systems undergoing FOPT. 



Further consideration concerns only the case of an itinerant system. 
We need to obtain the magnetic field dependence of chemical potential $\mu  = \muc(T,h)$ 
to keep the system in the PS regime at finite $h$. 
The condition that the system remains in PS region when a~magnetic field is applied is given by~Eq.~(\ref{eq:PS_condition}). 
Differentiating Eq.~(\ref{eq:PS_condition}) with respect to $h$ and taking into account the relation $\partial\Omega/\partial m = 0$ at $m = m_{\rm PM, FM}$, we get
\begin{multline}\label{eq:Omega1=Omega2}
	\frac{\partial\Omega}{\partial h}(T, \muc(T,h), h | m_{\rm PM}) + \frac{\partial\Omega}{\partial \mu}(T, \muc(T,h), h | m_{\rm PM})\left(\frac{d\mu}{dh}\right)_{\rm PS} \\ =	\frac{\partial\Omega}{\partial h}(T, \muc(T,h), h | m_{\rm FM}) + \frac{\partial\Omega}{\partial \mu}(T, \muc(T,h), h | m_{\rm FM})\left(\frac{d\mu}{dh}\right)_{\rm PS}.
\end{multline}
We obtain some analog of~the~Clausius–Clapeyron relation determining the chemical potential derivative in Eq.~(\ref{eq:chi_general}) in the PS state as
\begin{equation}\label{eq:eta_1st_order}
	\eta_{\rm PS} \equiv \left(\frac{d\mu}{dh}\right)_{{\rm PS}, h = 0} = -\frac{m_{\rm FM} - m_{\rm PM}}{n_{\rm FM} - n_{\rm PM}},
\end{equation}
instead of $\eta_{\rm n}$ for~single-phase case, see~Eq.~(\ref{eq:eta_def_n}). 
Substituting the Eq.~(\ref{eq:Landau_exp}) into the last equation, one obtains
\begin{equation}
	\eta_{\rm PS} =(1/\mf)(\dot{A}^\Omega_2 + \dot{A}^\Omega_4\mfmf + \dot{A}^\Omega_6 m^4_{\rm FM})^{-1}.
\end{equation}
We have common $\eta = \eta_{\rm PS}$ for both phases involved in PS differing by $m$ value only. 
Then, using the equation of state (\ref{eq:implicit_eq_state}), we get $\Phi$ \textit{phase} magnetic susceptibility $\chi^{\rm PS}_{\Phi} = dm_\Phi(T,\mu_{\rm PS}(T,h),h)/dh$
\begin{equation}\label{eq:chi_PS_general}
\chi^{\rm PS}_{\Phi}  = \bar\chi_{\Phi}\left(1 - \eta_{\rm PS} \left(\frac{\partial f}{\partial \mu}\right)_{\Phi}\right).
\end{equation}
So, we get directly
\begin{eqnarray}
	\chi^{\rm PS}_{\rm PM} &=& \bar\chi_{\rm PM},\\
	\chi^{\rm PS}_{\rm FM} &=& -\bar\chi_{\rm FM}\left(1 + \frac{2\mfmf(\dot{A}^\Omega_4 + 2\dot{A}^\Omega_6 \mfmf)}{\dot{A}^\Omega_2 + \dot{A}^\Omega_4\mfmf + \dot{A}^\Omega_6 m^4_{\rm FM}}\right).
\end{eqnarray}
For the vicinity of tricritical point $\mf$ is small 
then the second term in brackets in the last equation can be neglected and one get in the leading order with respect to $m$
\begin{equation}
	\chi^{\rm PS}_{\rm FM} \simeq -\frac1{8A^\Omega_2} < 0.
\end{equation}
One can see that 
in the vicinity of tricritical point $\chi^{\rm PS}_{\rm FM} $ is essentially negative (inverse sign with respect to Eq.~(\ref{eq:bar_chi_FM})) and equal by absolute value to $\bar\chi_{\rm FM}$ from Eq.~(\ref{eq:bar_chi_FM}). 

Above we discuss the~phase magnetic susceptibilities of PM and FM phases involved in PS. 
We have shown that these two phases behave differently under the change of the magnetic field.
There is significant contribution to magnetic susceptibility of FM phase involved in PS state that comes from a change in chemical potential as a function of the magnetic field (see Eqs.~\eqref{eq:eta_1st_order} and \eqref{eq:chi_PS_general}), while for PM phase there is no such contribution since $\left(\partial f/\partial \mu\right)_{\rm PM}=0$.
Let us emphasize here that total response for a system in the PS state should be obtained as a phase weighted average (see details in the model example below). 
To our knowledge this result is new, although some conceptually close results were obtained earlier for the~explanation of~martensitic phase transition temperature dependence on~a~magnetic field in~steels~\cite{Krivoglaz}.

\subsection{Entropy calculation}

We are mainly interested in the calculation of $(\Delta S)_{\rm env}$ which is dependent on the physical environment (we mark this by~the~index ``env'', (i) ``env''$=$ ``$n$'' and (ii) ``env'' $=$ ``PS'')  
\begin{equation}
	(\Delta S)_{\rm env} = \int_0^h dh' \left(\frac{dS(h')}{dh'}\right)_{\rm env},
\end{equation}
thus the sign of $(\Delta S)_{\rm env}$ is determined by the sign of $(dS/dh)_{\rm env}$ at 
small enough magnetic field. 

One can see that entropy definition $S(T, \mu | m) = -{\partial \Omega(T, \mu | m)}/{\partial T}$ does not depend on magnetic field $h$ \textit{explicitly}, but only through chemical potential $\mu$ and magnetization $m$. So we have
\begin{equation}\label{eq:dS_dh_general}
    \left(\frac{dS_\Phi}{dh}\right)_{\rm env} = \left(\frac{\partial S}{\partial m}\right)_{m=m_\Phi}\chi^{\rm env}_\Phi + \left(\frac{\partial S}{\partial \mu}\right)_{m = m_\Phi}\eta_{\rm env},
\end{equation}
where $\eta_{\rm env}$ is determined by~Eq.~\eqref{eq:eta_def}. 

Now we discuss $({dS}/{dh})_{\rm env}$ expressions for both single-phase and phase-separated cases and compare them with each other. For the single-phase case one gets 
\begin{equation}
	 \left(\frac{dS_\Phi}{dh}\right)_n = \chi^n_\Phi\left(\frac{dS_\Phi}{dm}\right)_n,
\end{equation}
where 
\begin{equation}
\left(\frac{\partial S}{\partial m}\right)_n = \left(\frac{\partial S}{\partial m} - \frac{\partial S}{\partial \mu}\frac{{\partial \mathfrak{n}}/{\partial m}}{{\partial \mathfrak{n}}/{\partial \mu}}\right)_{m = m_\Phi},
\end{equation}
with $\chi^n_\Phi$ defined by~Eq.~(\ref{eq:chi_n}). 

Consider how the presence of PS and the negative sign of the magnetic response can affect the entropy change when the magnetic field is applied. Let $S_{\Phi}(T,h) = S (T, \mu_{\rm PS}(T,h) | m_{\Phi}\left(T,\mu_{\rm PS}(T,h),h\right))$ be the entropy of~the~phase $\Phi$ involved in~PS. 

From Eq.~\eqref{eq:dS_dh_general}, one gets the entropy change derivative for the phase-separated case for the phase $\Phi$ involved in PS
\begin{equation}\label{eq:dSdh_PS_state}
   \left(\frac{dS_{\Phi}}{dh}\right)_{\rm PS, \Phi} = \left(\frac{\partial S}{\partial m}\right)_{m=m_\Phi}\chi^{\rm PS}_{\Phi} + \left(\frac{\partial S}{\partial \mu}\right)_{m=m_\Phi}\eta_{\rm PS},
\end{equation}
where $\eta_{\rm PS}$ is determined by~Eq.~(\ref{eq:eta_1st_order}). 
Here it is taken into account that when the magnetic field is applied, not only the magnetization of the ferromagnetic phase $m$ changes, but also $\mu$ does. 
In the vicinity of tricritical point one can expect that the first ($\chi$) term dominates over the second ($\eta$) term; 
this fact was verified within numerical calculations, see~Sec.~\ref{sec:calculation_example}. 

Thus, we have
\begin{equation}
	\left(\frac{dS_{\Phi}}{dh}\right)_{\rm PS} = \frac{\chi^{\rm PS}_{\Phi}}{\chi^n_{\Phi}}\left(
 \left(\frac{dS_{\Phi}}{dh}\right)_n + \left(\frac{\partial S}{\partial \mu}\right)_{m=m_{\Phi}}\left(\frac{\chi^n_{\Phi}}{\chi^{\rm PS_{\Phi}}}\eta_{\rm PS} - \eta_{n}\right)
 \right).
\end{equation}  
In FM phase, we have two contributions to the infinitesimal entropy change: the $\chi$ term is opposite to the usual sign due to the negative phase susceptibility  and $\eta$ term has an~uncertain sign. We have verified that entropy derivatives in the equation above have comparable values and so we need to account for both of them.

We state that the participation of FM phase in PS substantially can generally change both magnetic and entropy response of the FM phase. 
Furthermore, PS presence can switch MCE from direct one outside the PS region to inverse one within the PS region. 
For the paramagnetic phase involved in PS, the phase magnetic susceptibility is positive, see~Eq.~(\ref{eq:dmdh}). 
Thus, in the phase-separated (mixed) state, the magnetic response of the paramagnetic and ferromagnetic phases are opposite in sign, which can lead to the fact that the ferromagnetic phase contributes an inverse sign to the isothermal change of the entropy $\Delta S$ when the magnetic field is applied. 

\section{Entropy change for a  first-order FM-PM transition induced by giant an Hove singularity: the case of fcc lattice}\label{sec:calculation_example} 

In this Section we consider MCE for the~case of non-degenerate Hubbard model for~the~face-centered cubic lattice having giant van Hove singularity at the band bottom. 
Generalized Landau theory coefficients $A^\Omega_\alpha$ (see previous Section) and the  isothermal change of the entropy $\Delta S(T)$ 
are calculated and analyzed in detail. 
The Hamiltonian reads
\begin{equation}\label{eq:Hamiltonian}
    \mathcal{H}=\sum_{ij\sigma}t_{ij}c^\dag_{i\sigma}c^{}_{j\sigma} + U \sum_i n_{i\uparrow}n_{i\downarrow}-h\sum_{i\sigma}\gamma_\sigma n_{i\sigma},
\end{equation}
where $t_{ij}$ is non-zero and equal to $t$~($t'$), when sites  $i$ and $j$ are nearest~(next-nearest) neighbours, $c^\dag_{i\sigma}/c_{i\sigma}$ is a~creation/annihilation electron operator,  and $n_{i\sigma} = c^\dag_{i\sigma}c_{i\sigma}$ is an~operator of number of electrons at~the~site~$i$ with spin projection $\sigma=\uparrow,\downarrow$, $\gamma_{\uparrow(\downarrow)}=+1(-1)$, $U$~is matrix element of local Coulomb interaction. 

\begin{figure}	
  \includegraphics[angle=-90,clip=true,width=0.79\textwidth]{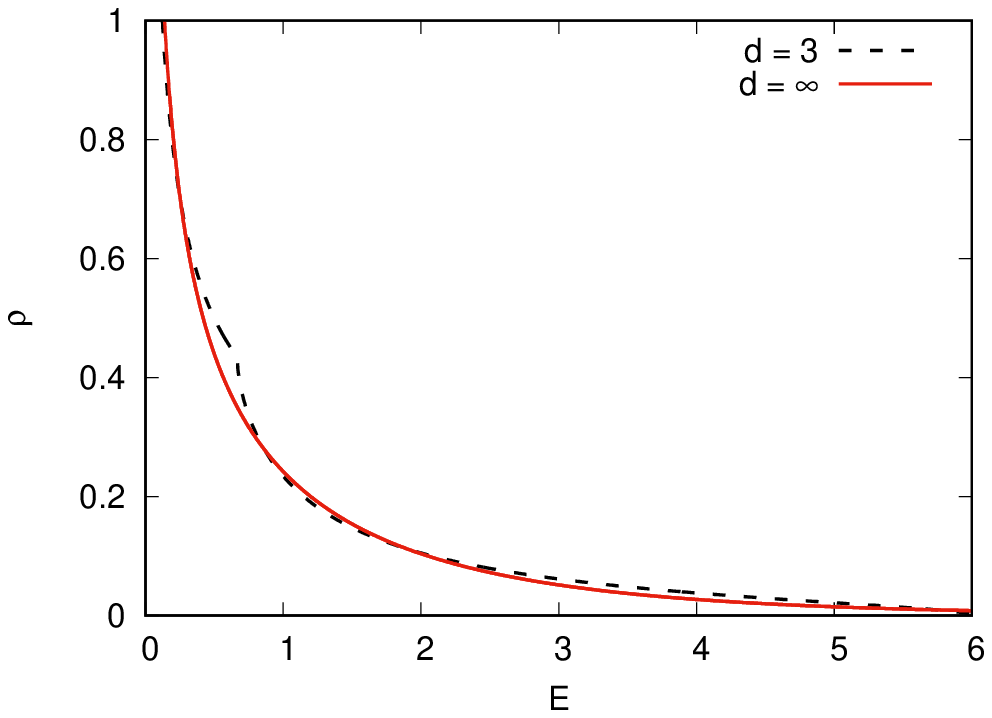}
\caption{Dimensionless density of electron states for a next-nearest neighbor infinite-dimensional ($d = \infty$) $\rho^\infty_{\rm fcc}(E)$ and three-dimensional ($d = 3$) fcc lattice with $t' = -t/2$, see Eq.~(\ref{eq:DOS_fcc}) and Refs.~\onlinecite{Ulmke,2022:Igoshev_DOS_fcc}. The dimensionless energy variable $E = 1 + \epsilon/(3t)$ is counted from the band bottom. 
	\label{fig:DOS}
	}	 
\end{figure}

To obtain giant van Hove singularity at the band bottom, we consider the case $t' = -t/2$, so
the density of states $\tilde{\rho}_{\rm fcc}(\epsilon) = (1/N)\sum_{\mathbf{k}}\delta\left(\epsilon - \epsilon^{\rm fcc}_\mathbf{k}\right)$ for bare fcc electron spectrum $\epsilon^{\rm fcc}_\mathbf{k} = (1/N)\sum_{ij}t_{ij}\exp[\I\mathbf{k}(\mathbf{R}_i - \mathbf{R}_j)]$, where $N$ is a site number, can be written
\begin{equation}\label{eq:dos_fcc_tau=-1/2}
    \tilde{\rho}_{\rm fcc}(\epsilon) = \frac1{t}\sqrt{\frac{2}{3E(\epsilon)}}\rho_{\rm sc}\left(\sqrt{6E(\epsilon)}\right)
\end{equation}
through dimensionless energy variable $E(\epsilon) = 1 + \epsilon/(3t)$,
where $\rho_{\rm sc}\left(\varepsilon\right)$
is dimensionless density of state~(DOS) of~simple cubic lattice spectrum $\epsilon^{\rm sc}_{\mathbf{k}}$ within nearest neighbor approximation with unit hopping integral,  
see Refs.~\onlinecite{Vollhardt,Ulmke,2022:Igoshev_DOS_fcc}. 
This expression 
has giant one-dimension-like (square-root) van Hove singularity at the band bottom, $\epsilon = -3t$ or $E = 0$. 
To simplify calculations we replace $\rho_{\rm sc}(\varepsilon)$ by~DOS~of~infinite-dimensional hypercubic lattice~(hc) spectrum $\epsilon^{\rm hc}_{\mathbf{k}}$ DOS in the nearest neighbor approximation   
\begin{equation}\label{eq:dos_sc_d=inf}
\rho^\infty_{\rm hc}(\varepsilon) = \frac1{\sqrt{12\pi}}\exp[-\varepsilon^2/12]
\end{equation}
with hopping integral chosen in such a way to provide 
$\sum'_\mathbf{k}\left({\epsilon^{\rm sc}_{\mathbf{k}}}\right)^2=\sum'_\mathbf{k}\left({\epsilon^{\rm hc}_{\mathbf{k}}}\right)^2$, where the notation $\sum'_\mathbf{k}\equiv (1/N)\sum_\mathbf{k}$ is used.  
From~Eqs.~(\ref{eq:dos_fcc_tau=-1/2})~and~(\ref{eq:dos_sc_d=inf}) we get
the expression for dimensionless DOS in terms of $E$ variable 
\begin{equation}\label{eq:DOS_fcc}
\rho_{\rm fcc}^\infty(E) = \frac{\exp(-E/2)}{\sqrt{2\pi E}}
\end{equation} 
which only weakly deviates from three-dimensional case fcc lattice DOS, $\rho_{\rm fcc}(E)$ written through $E$ variable, in the vicinity of the band bottom, see~Fig.~\ref{fig:DOS}. 
Below we take $3t$ as energy unit. 

Further the Hubbard model is treated within the mean-field (Hartree-Fock) approximation~(HFA) under the assumption that giant van Hove singularity at~the~band bottom induces the~ferromagnetic ordering. 
In terms of functional renormalization group method, the application of HFA with renormalized (screened) interaction parameter $U$ here is justified by the dominance of the crossed particle-hole electron-electron interaction channel at low temperature~\cite{FRG_review,2023:Igoshev_fcc_fRG}. 
Due to the presence of a van Hove singularity in the vicinity of the Fermi level   the Stoner criterion may be fulfilled with small enough $U$ parameter. 

A magnetization $m$ and a chemical potential $\mu$ are determined by following HFA equations 
\begin{equation}
	\label{eq:n-Stoner}
	n = \sum_{\kk\sigma}' f_{\mu_0}(\varepsilon^{\rm fcc}_{\kk\sigma}),
\end{equation}
and 
\begin{equation}
	\label{eq:m-Stoner}
	m = \sum_{\kk\sigma}' \gamma_\sigma f_{\mu_0}(\varepsilon^{\rm fcc}_{\kk\sigma}),
\end{equation}
where $f_{\mu}(\varepsilon) = [\exp[(\varepsilon - \mu)/T] + 1]^{-1}$ is the Fermi function with HFA electronic spectrum $\varepsilon^{\rm fcc}_{\kk\sigma} = \epsilon^{\rm fcc}_{\kk} - \gamma_\sigma(Um/2 + h)$, 
$\mu_0 = \mu - Un/2$. 

At first, we discuss the~conventional approach, based on the use of $n$ as a~main variable, allowing us to describe the properties of paramagnetic and ferromagnetic phases for single-phase state. In this case, the thermodynamical potential is a~free energy per site~\cite{Moriya} 
\begin{equation}\label{eq:F_HFA}
F_{\rm HFA}(T,n,h|m) = \Omega_{\rm HFA}(T,\mu,h|m) + \mu n,
\end{equation}
where $\mu$ is determined by $\partial\Omega/\partial\mu = 0$ and HFA grand potential of a single-phase state reads 
\begin{multline}\label{eq:Omega_HFA}
\Omega_{\rm HFA}(T,\mu,h|m) = \Omega_0(T, \mu_0(T,n_{\rm HFA}(T,\mu,m),m),Um/2 + h) 
\\
- (U/4)(n^2_{\rm HFA}(T,\mu,m) - m^2),  
\end{multline}
with grand potential of band (free) electrons  
\begin{equation}
\Omega_0(T,\mu,h) =
2\mu 
- {T}\sum_{\mathbf{k}\sigma}'\ln\left[2\cosh\frac{\epsilon^{\rm fcc}_{\mathbf{k}}  - \gamma_\sigma h - \mu}{2T}\right],
\end{equation}
where $n = n_{\rm HFA}(T,\mu,m)$ is a~solution of~the~equation
\begin{equation}\label{eq:n_HFA_def}
\mu_0(T,n,m) = \mu - Un/2.
\end{equation}
Here,  auxiliary chemical potential $\mu = \mu_0(T,n,m)$  is required to setup the magnetization $m$ and band filling $n$ values for free electron gas.

The expansion of free energy defined by~Eq.~(\ref{eq:F_HFA}) is 
\begin{equation}\label{eq:F_exp}
F_{\rm HFA}(T,n,h|m) = A^F_0(T,n) + A^F_2(T,n)m^2 + A^F_4(T,n)m^4 +  A^F_6(T,n)m^6 - hm,
\end{equation}
where $A^F_0(T,n)$ is a free energy at $m = 0$. 
Also,
\begin{eqnarray}
\label{eq:A2_F}
A^F_2(T,n) &=& \frac14(\Pi^{-1} - U), \\
\label{eq:A4_F}
A^F_4(T,n) &=& \frac1{64\Pi^5}\left({{\Pi'}^2} -  \frac13{\Pi\Pi''}\right), \\
\label{eq:A6_F}
A^F_6(T,n) &=& \dfrac{1}{1536 \Pi^7}\left(\frac{7{\Pi'}^2}{\Pi^2} \left({\Pi'}^2 - \Pi\Pi''\right) + \dfrac{2}{3}{\Pi''} ^2 + \Pi'\Pi^{'''} - \dfrac{1}{15}{\Pi\Pi''''}\right),
\end{eqnarray}
where $\Pi = \Pi(T,E_{\rm F})$ with Fermi level corresponding to given band filling~$n$, where
\begin{equation}
\Pi(T, E_{\rm F}) = -\sum_{\mathbf{k}}' f'_{E_{\rm F}}(\epsilon^{\rm fcc}_{\mathbf{k}})  
\end{equation}
is up to constant factor magnetic susceptibility in the paramagnetic phase. Prime above $\Pi$ denotes the~derivative with~respect to~$E_{\rm F}$ . 

\begin{figure}
        \includegraphics[angle=-90,clip=true,width=0.69\textwidth]{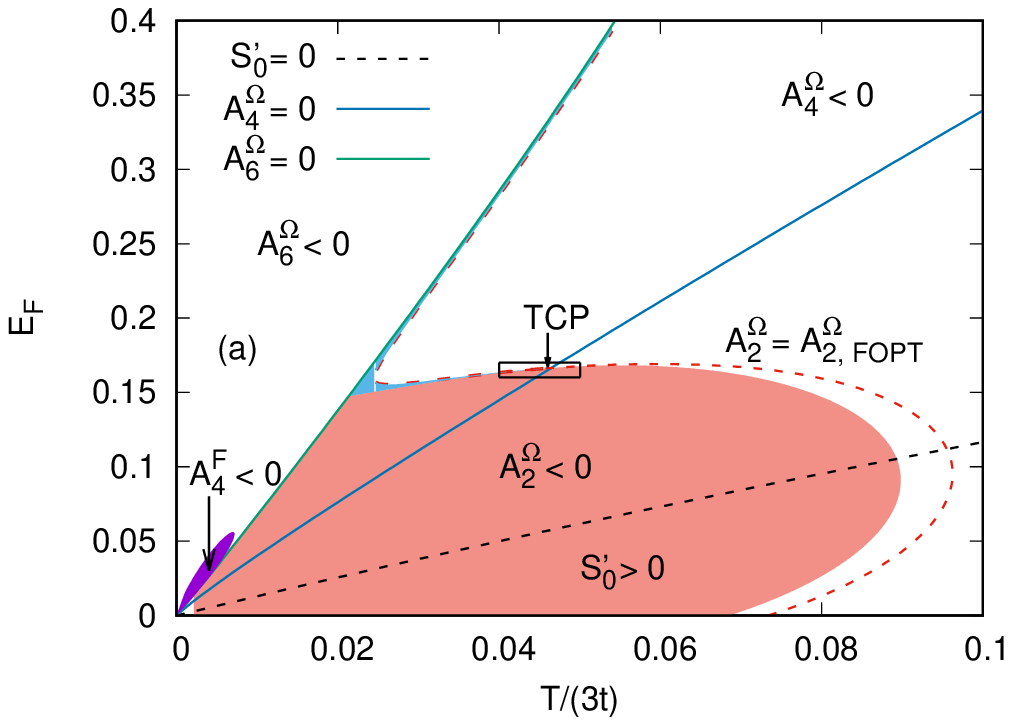}     
        \includegraphics[angle=-90,clip=true,width=0.69\textwidth]{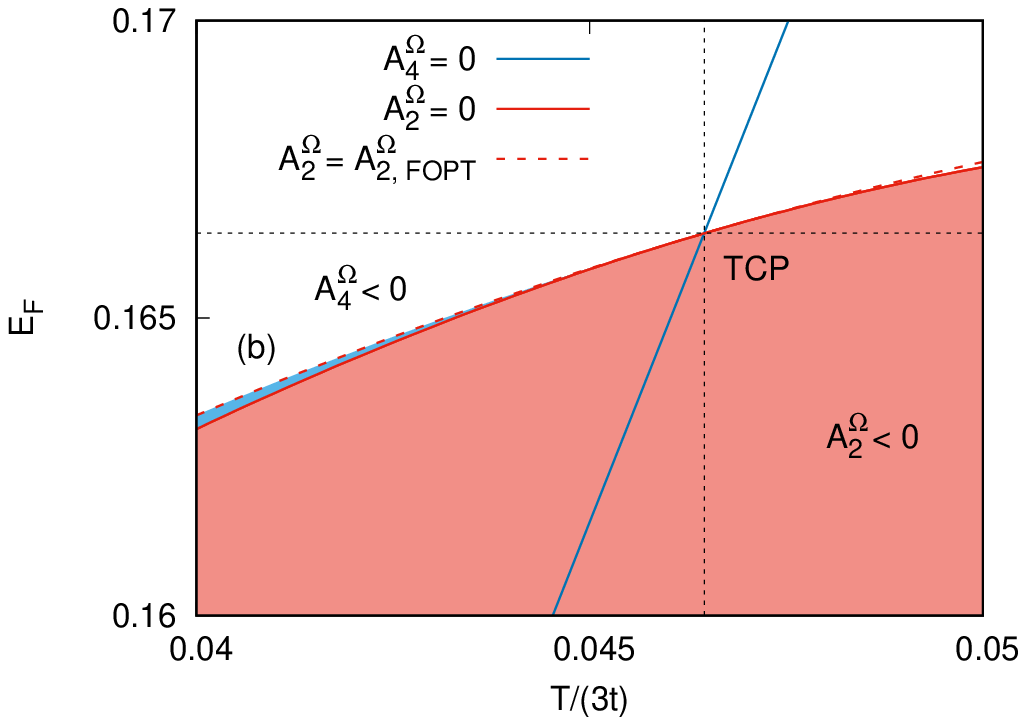}     
\caption{
(a)~Phase diagram at $U = 3t$ in terms of~$T$ and dimensionless Fermi level $E_{\rm F}$ calculated within the generalized Landau theory with coefficients $A^{\Omega}_2$, $A^{\Omega}_4$, $A^{\Omega}_6$, see Eqs.~(\ref{eq:A2}-\ref{eq:A6}) in~Sec.~\ref{sec:Landau_theory}. $A^{\Omega}_2 = 0$ corresponds to SOPT line in the region $A^{\Omega}_4, A^{\Omega}_6 > 0$, red dashed line corresponds to $A^{\Omega}_2 = A^{\Omega}_{2,\rm FOPT}$. This line corresponds to FOPT line inside $A^{\Omega}_4 < 0$ region. Dashed black line corresponds to $S_0'(T,E_{\rm F}) = 0$, see~Eq.~\eqref{eq:S0_def}.  Violet region indicates the region $A^F_4(T,E_{\rm F}) < 0$ inside.  Tricritical point (TCP) position is shown by an arrow. Region $A^{\Omega}_2 < 0$ is filled by red color, the region $0 < A^{\Omega}_2 < A^{\Omega}_{2,\rm FOPT}$, $A^{\Omega}_4 < 0$, $A^{\Omega}_6 > 0$ is filled blue color.  Black rectangular shows the region of vicinity of TCP shown in Fig.~\ref{fig:Landau_PD}(b). 
(b) The same as (a)~in~the~vicinity of~TCP. Dotted lines show the position of tricrical point.
}
\label{fig:Landau_PD}
\end{figure}

The calculation of the dependence $A^F_4$ on $T$ and $E_{\rm F}$ using Eq.~\eqref{eq:A4_F} allows us to~analyze the applicability of conventional approach. 
In the case $A^F_4 > 0$ conventional Landau theory yields the phase transition line $A^F_2 = 0$, separating FM region at $A^F_2 < 0$, or equivalently
\begin{equation}
	\label{eq:Stoner_cr}
	U\Pi(T, E_{\rm F}) > 1,
\end{equation}
from paramagnetic region at $A^F_2 > 0$. In the limit $T\rightarrow0$, $\Pi(T, E_{\rm F}) \rightarrow \rho(E_{\rm F})$ which transforms~Eq.~\eqref{eq:Stoner_cr} into the~conventional Stoner criterion. 
Numerical calculation at $U = 3t$ shows that $A^F_4$ is negative in the small region in the vicinity of van Hove singularity position at small temperatures, see Fig.~\ref{fig:Landau_PD}. 
We find that singular behavior of $\rho^\infty_{\rm fcc}(E)$ near the band bottom can change the sign of $A^F_4$ and this makes the conventional Landau theory inapplicable. In this situation it is necessary to~consider accurately the possibility of FOPT FM-PM. 

For itinerant electron system the FOPT results in electronic phase separation which is convenient to investigate using $\mu$ as main variable and $\Omega_{\rm HFA}(T, \mu, h | m)$ instead of $F_{\rm HFA}(T, n, h | m)$ (see discussion in~Refs.~\onlinecite{IgoshevJMMM,2010:Igoshev,2015:Igoshev_JPCM}). 
Such transition is accompanied not only by a jump of order parameter but also by a~jump of band filling. 

Instead of expansion (\ref{eq:F_exp})  
one can use the expansion of $\Omega_{\rm HFA}$, see Eq.~\eqref{eq:Omega_HFA}, cf.~Sec.~\ref{sec:Landau_theory}, with explicit expressions for coefficients 
\begin{eqnarray}
\label{eq:A2}
A^\Omega_2(T,\mu) &=& \frac1{4}(\Pi^{-1} - U), \\
\label{eq:A4}
A^\Omega_4(T,\mu) &=& \frac1{{192f\Pi^4}}(3U{\Pi'}^2 - f\Pi''), \\
\label{eq:A6}
A^\Omega_6(T,\mu) &=& \frac1{720}\frac1{8f^3\Pi^7} \left(
9(11f-1)U^2{\Pi'}^4 - 3f(31f - 11)U\Pi''{\Pi'_0}^2
\right.
\\
&&
\left.
+ 2f^2(2f + 3){\Pi''}^2 
+ 15f^2(f-1)\Pi'''\Pi'
-f^3\Pi''''\Pi
\right),
\end{eqnarray}
where $f = 1 + U\Pi$, arguments $T, \mu$ of $\Pi$ are omitted for brevity. 
Note that $A^\Omega_2(T,\mu) = A^F_2(T,n)$ with $n$ and $\mu$ related by Eq.~\eqref{eq:n-Stoner},  but $A^\Omega_4 \ne A^F_4$, $A^\Omega_6 \ne A^F_6$. 
The derivation of Eqs.~(\ref{eq:A2}-\ref{eq:A6}) will be published later~\cite{Igoshev:unpub}. 

The calculation of $A^\Omega_\alpha(T,\mu)$, $\alpha = 2,4,6$ using Eqs.~(\ref{eq:A2}-\ref{eq:A6}) allows us to calculate the phase diagram within the~generalized Landau theory, see~Fig.~\ref{fig:Landau_PD}. Accordingly to~the~discussion of~Sec.~\ref{sec:Landau_theory}, in the case $A^\Omega_6>0$ FM region corresponds to $A^\Omega_2 < 0$ at $A^\Omega_4 > 0$ and $A^\Omega_2 < A^\Omega_{2,\rm FOPT}$ for $A^\Omega_4 < 0$. Otherwise PM phase is established. 
In the case $A^\Omega_4 < 0$,  
the equation $A^\Omega_2 = A^\Omega_{2,\rm FOPT}$ yields FOPT line
with corresponding PS region (see below); in the case $A^\Omega_4 > 0$ the equation $A^\Omega_2 = 0$ yields SOPT line. 
For the case $A^\Omega_6 < 0$, the consideration of higher-order terms of generalized Landau expansion is necessary. 
While using $\Omega$ expansion allows to make qualitative conclusions about phase diagram,  
it is very inconvenient for detailed calculations since necessary number of terms in the expansion series~\eqref{eq:Landau_exp} to be held is \textit{a priori} unknown. To this end a better way is straightforward minimization of $\Omega_{\rm HFA}$ given by Eq.~(\ref{eq:Omega_HFA}) by solving extremum equations~(\ref{eq:n-Stoner}-\ref{eq:m-Stoner}). 

For definiteness we suppose that two possible phases are characterized by $n_{\rm FM} (\mu, T, h), m_{\rm FM} (T, \mu, h)$, and $n_{\rm PM} (T, \mu, h), m_{\rm PM} (T, \mu, h)$. 
Each of these sets of parameters corresponds to a~local minimum of $\Omega_{\rm HFA}(T,\mu,h|m)$. 
In our consideration chemical potential $\mu$ controls the position of global minimum of $\Omega_{\rm HFA}(T,\mu,h|m)$.
So one can assume that there exists some critical value $\mu=\mu_{\rm PS}$ so that
$\Omega_{\rm HFA}(\mu, T, h | m_{\rm FM}) <\Omega_{\rm HFA} (\mu, T, h | m_{\rm PM})$ for $\mu <\mu_{\rm PS}(T, h) $ and $\Omega_{\rm HFA}(\mu, T | m_{\rm FM})> \Omega_{\rm HFA} (\mu, T, h | m_{\rm FM}) $ for $\mu> \mu_{\rm PS} (T, h)$ (arguments $\mu$, $T$ and $h$ of $m_{{\rm FM},{\rm PM}}$ are omitted for brevity), cf.~$\mu_{\rm PS}$ definition in~Sec.~\ref{sec:Landau_theory}. 
Thus, as $\mu$ increases, both $m$ and $n$ jump at $\mu = \mu_{\rm PS} $, and a state with $n$ satisfying $n_{\rm FM} < n <n_{\rm PM} $ corresponds to phase-separated state. 
The case of the second-order magnetic phase transition is reproduced within this scheme when $m_{\rm FM} = m_{\rm PM}$ (and therefore $n_{\rm FM} = n_{\rm PM}$): in this case PS is absent and treatments using $F_{\rm HFA}$ and $\Omega_{\rm HFA}$ potentials are equivalent. 
An application of $F_{\rm HFA}$-based scheme in the case $n_{\rm FM} < n < n_{\rm PM}$ results in thermodynamically unstable single-phase state. 
This is fully analogous to liquid-gas first-order phase transition under pressure in molecular physics, where $(-\mu)$ ($n$) plays the role of pressure (density). 
It should be noted that we assume that the system is always in the state corresponding to~a~minimum of the grand potential $\Omega_{\rm HFA}$ or $\Omega$. 
Therefore metastable states producing hysteresis effects are beyond our~consideration. 

\begin{figure}
\includegraphics[angle=-90,clip=true,width=0.89\textwidth]{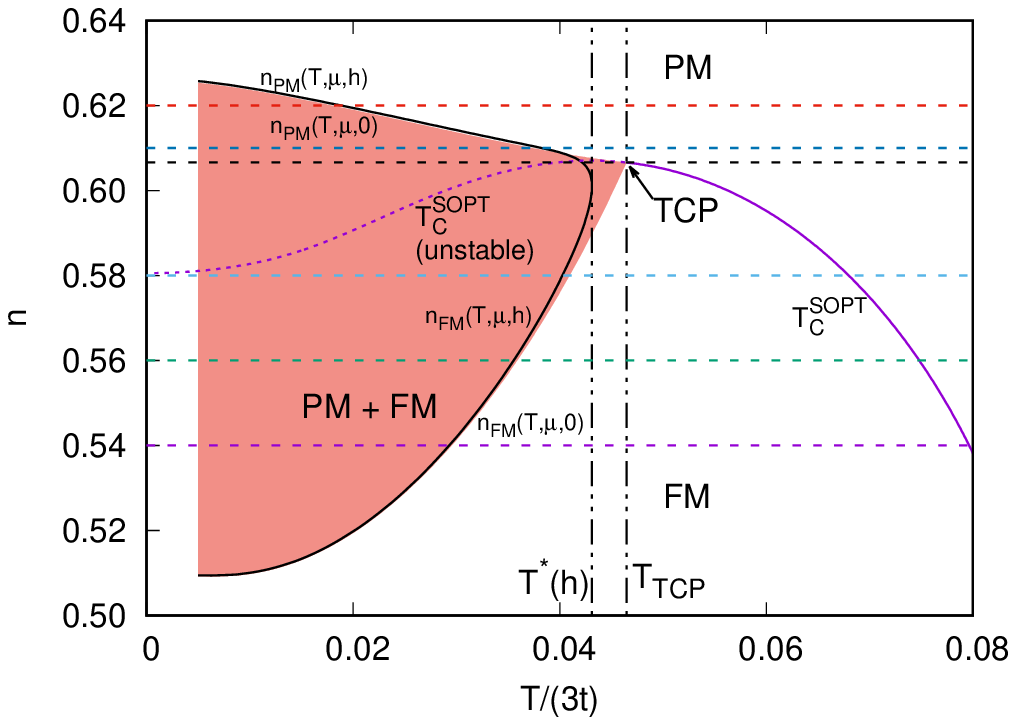}
\caption{$T-n$ phase diagram of the Hubbard model for an infinite-dimensional fcc lattice with $U = 3t$.
PS region FM~+~PM at $h = 0$ is filled by red color and bounded by a line $T = T_{\rm PS}(n, 0)$, black solid line indicates the corresponding~PS boundary at~finite~$h = 3\cdot10^{-4}t$ and bounded by a line $T = T_{\rm PS}(n, h)$. 
Purple line is PM-FM phase boundary $T = T^{\rm SOPT}_{\rm C}$ obtained from the Stoner criterion $U\Pi(T, E_{\rm F}) = 1$. 
Its solid part corresponds to second-order phase FM-PM transition,  
whereas dotted part has no sense (unstable) since it falls into the PS region. 
The vertical dashed lines show the temperature boundaries of the phase separation in zero $ T_{\rm TCP}$~(right) and finite $T^\ast(h)$~(left) magnetic field. 
Colors of horizontal dashed lines correspond to the values of $n$ presented in Fig.~\ref{fig:T_vs_S.gallery}.}
\label{fig:T_vs_n}  
\end{figure}

We have chosen $U = 3t$, which is reasonable estimation value for screened Coulomb interaction parameter and small enough to justify HFA, and have calculated the $T-n$ phase diagram by direct comparing $\Omega_{\rm HFA}$ for ferromagnetic and paramagnetic solutions of HFA equations. 
Using rather dense grid by $\mu$ and $T$ variables allows to localize the position $\mu = \mu_{\rm PS}$ where $\Omega_{\rm HFA}(T,\mu, h| m_{\rm FM})  = \Omega_{\rm HFA}(T,\mu, h| m_{\rm PM})$ for numerically calculated $\Omega_{\rm HFA}$. 
The Fig.~\ref{fig:T_vs_n} presents the  phase diagram   including FM$+$PM phase separation region due to first-order transition at low temperatures for $ n \simeq 0.6$ which provides the closeness of Fermi level to the band bottom (van Hove singularity). 
We have verified that the change of $U$ yields similar results. 
While in terms of $T-E_{\rm F}$, see~Fig.~\ref{fig:Landau_PD}, FOPT line is very close to SOPT line, corresponding to FOPT line PS region in terms of $T-n$ is rather wide. 
PS boundaries are shown in two variants: in a zero and finite magnetic field $h = 3\cdot10^{-4}t$. 
PS region is bounded by $n = n_{\rm PM}(T, h)$ below and $n = n_{\rm FM}(T, h)$ above. These curves merge and PS disappears at $T = T^{\ast}(h)$ which corresponds to the point $ (T^{\ast}(h), n^\ast(h))$, whose position depends on $h$. 
At $h = 0$, the point $(T^{\ast}(h), n^\ast(h))$ becomes TCP: $T_{\rm TCP} \equiv T^\ast (h = 0) = 0.04645\cdot 3t$, $n^\ast_0\equiv n^\ast (h = 0) = 0.6066$. 
At this point, the~phase transition order change occurs at $h = 0$: for $ T> T_{\rm TCP}$, the phase transition is of second-order and the transition point is determined by the condition $U\Pi (T, E_{\rm F}) = 1$ which determines SOPT Curie temperature $T^{\rm SOPT}_{\rm C}$; for $ T<T_{\rm TCP}$ the curve determined by this condition is inside the PS region and loses its meaning. 
We denote the temperature boundary of the PS region for both lines $n_{\rm FM,PM}(T, h)$ as $T_{\rm PS}(n, h)$. 
From Fig.~\ref{fig:T_vs_n} it can be seen that when $n\simeq n^\ast_0 $ the difference $T_{\rm PS}(n, h) $ from $T_{\rm PS} (n, h = 0) $ can be significant. 

Below we study $\Delta S(T)$ for $T$ near temperature boundary $T_{\rm PS}(n,h)$. 
Similar investigations of $\Delta S(T)$ for Bethe lattice (second order phase transition) and square lattice (first order phase transition) are presented in Refs.~\onlinecite{2017:Igoshev,2021:Ivchenko,2023:Igoshev_MCE_PS_square_lattice}. 
Here we calculate the entropy per site (in units of Boltzmann constant $k_{\rm B}$) within the HFA as
\begin{equation}\label{eq:S_HFA}
	S(\mu, T, h) = \sum_{\sigma} S_0(T, E^{\rm HFA}_{\rm F,\sigma}(\mu, n, m)),
\end{equation}
where 
\begin{equation}\label{eq:EF_sigma}
    E^{\rm HFA}_{\rm F,\sigma}(\mu, n, m) = \mu - Un/2 + \gamma_\sigma (Um/2 + h)
\end{equation}
being the effective Fermi level of $\sigma$ spin subband for HFA solution, see~Eqs.~(\ref{eq:n-Stoner}--\ref{eq:m-Stoner}). 
An~entropy per site and one spin subband  for non-interacting electrons with spectrum $\epsilon^{\rm fcc}_{\kk}$ and Fermi level $E_{\rm F}$ is 
\begin{equation}\label{eq:S0_def}
	S_0(T, E_{\rm F}) = \sum'_{\kk\sigma} \kappa\left(\frac{\epsilon^{\rm fcc}_{\kk} - E_{\rm F}}{T}\right)
\end{equation}
with an~auxiliary function
\begin{equation}\label{eq:kappa_def}
	\kappa(x) = \ln\left[2\cosh(x/2)\right] - (x/2)\tanh(x/2).
\end{equation}

To calculate volume-averaged thermodynamical quantities such as $m$, $S$ for the case of phase-separated state one needs to introduce the phase volume fractions $0< x_{\rm FM}, x_{\rm PM} < 1$ for given $n$ 
by the equation
\begin{equation}
	\label{eq:n_x}
	n = x_{\rm FM}n_{\rm FM}(T,\mu_{\rm PS}(T, h),h) + x_{\rm PM}n_{\rm PM}(T,\mu_{\rm PS}(T, h),h),	
\end{equation}
together with 
$x_{\rm FM} + x_{\rm PM} = 1$.	
Then any extensive quantity is just a corresponding phase-weighted average. 
For example, total magnetization and entropy then are
\begin{equation}
	\label{eq:m_x}
	m = x_{\rm FM}m_{\rm FM}(T,\mu_{\rm PS}(T, h),h) + x_{\rm PM}m_{\rm PM}(T,\mu_{\rm PS}(T, h),h),
\end{equation}
\begin{equation}
	\label{eq:S_x}
	S = x_{\rm FM}S_{\rm FM}(T,\mu_{\rm PS}(T, h),h) + x_{\rm PM}S_{\rm PM}(T,\mu_{\rm PS}(T, h),h),
\end{equation}
where $m_{\rm FM,PM}$~($S_{\rm FM,PM}$) is the phase magnetization~(entropy) for both phases involved in~PS.

\begin{figure}
\includegraphics[angle=-90,width=0.79\textwidth]{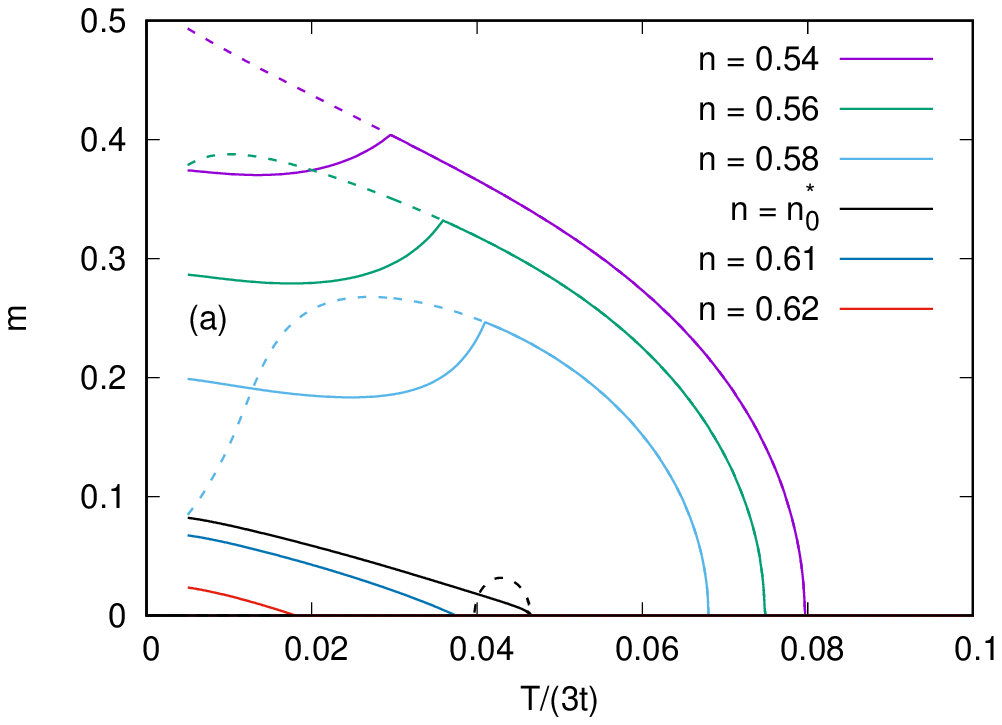}
\includegraphics[angle=-90,width=0.79\textwidth]{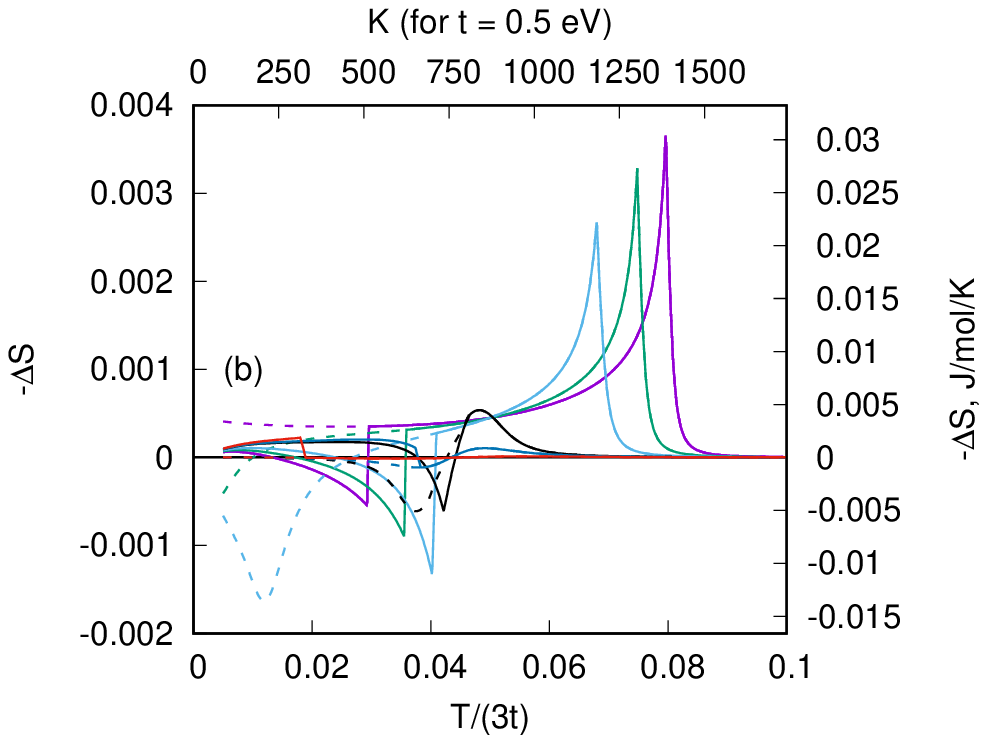}
	\caption{
 Temperature dependence of (a)~$m$, (b)~$\Delta S$ for fillings in the vicinity of $n = n^*_0 = 0.06066$ in the $[0.54,0.62]$ filling interval. $U = 3t, h = 3\cdot10^{-4}t$. 
 Particular filling is shown by color, see the legend of Fig.~\ref{fig:T_vs_S.gallery}. 
 Dashed lines show the result without PS formation.  For the sake of clarity on panel (b) we show $\Delta S$ in units of J/mol/K and, for $t=0.5$~eV value, temperature in units of K on additional axes. 
}
\label{fig:T_vs_S.gallery}  
\end{figure}

Figure~\ref{fig:T_vs_S.gallery} shows the temperature dependence of total  $m$ and $\Delta S$, see~Eqs.~(\ref{eq:m_x})~and~(\ref{eq:S_x}), for the case $U = 3t$ at various band fillings from the interval of PS existence $[0.50,0.63]$~(these values are shown by horizontal lines on $T-n$ phase diagram, see Fig.~\ref{fig:T_vs_n}). To illustrate the PS effect on $m$ and $\Delta S$ in~Fig.~\ref{fig:T_vs_S.gallery} the temperature dependence of these quantities is also calculated within the assumption that the system is a single-phase one. 

To clarify our results we convert them into commonly used units choosing
reasonable value for the nearest neighbor hopping integral $t = 0.5$~eV, so $U = 1.5$~eV, $T_{\rm TCP}\sim 0.04\cdot3t\sim 700$~K, $h = 3\cdot10^{-4}t$ corresponds to magnetic field $H = h/\mu_{\rm B}=26$~kOe ($\mu_{\rm B} = 5.8\cdot10^{-9}$~eV/Oe). To make it more accessible and convincing for experimental audience in~Fig.~\ref{fig:T_vs_S.gallery} we show $\Delta S$ in units of J/mol/K and, for the chosen $t$ value, temperature in units of K. (However, it is known that HFA approximation substantially overestimates critical temperature values~\cite{Moriya}.) Let us note, that numbers obtained have reasonable values.

The temperature behavior of $m$ and $\Delta S$ dramatically depends on~$n$~value. 
There is substantial effect of PS on $m$ temperature dependence: at the $T = T_{\rm PS} (n, 0)$ $m$ dependence exhibit a valuable kink. 
The height of $m(T)$ kink decreases when $n$ increases towards~$n^\ast_0$.  
For the calculation not taking into account PS formation,  $m(T)$ dependence has no kinks, instead there is pyromagnetic behavior, which is an artifact of~an~inconsistent treatment. 
When $T$ approaches finite-field PS boundary $T_{\rm PS}(n, h) $ with $ n <n^\ast_0$ $\Delta S $ has a sharp peak and is positive (inverse MCE). With further increase in temperature in a narrow interval $[T_{\rm PS} (n, h), T_{\rm PS} (n, 0)]$ $\Delta S$ is almost linearly dependent on temperature, rapidly changing from~positive value at the top of the peak to~moderate negative value. 
With a further increase in $T$, a typical negative peak for the second-kind transition with $\Delta S < 0$  at $T = T^{\rm SOPT}_{\rm C}$ takes place. 

For $n> n^\ast_0$ there is a weak feature (kink) of the dependence $\Delta S$ associated with leaving the PS region. $\Delta S$ peak height increases as $n$ approaches $n^\ast_0$, but there is no second-order transition.

For $n\approx n^\ast_0$, the behavior of $\Delta S $ is the most interesting. 
The peculiarity is formed by an inverse MCE peak at the point $T = T_{\rm PS}(n^\ast(h),h)$  and by a another  direct MCE peak-like feature \textit{not corresponding} to SOPT in the vicinity to $T = T_{\rm TCP}$, between which there is a~linear growth interval $ [T^\ast(h), T_{\rm TCP}]$~(for $ n\ne n^\ast_0$ this temperature range has a smaller width). 

The change of PS boundaries 
$n_{\rm FM}$, $n_{\rm PM}$ under the magnetic field applying is maximal at $ T = T_{\rm TCP}$. 
\begin{figure}[ht]
 \includegraphics[angle=-90,clip=true,width=0.79\textwidth]{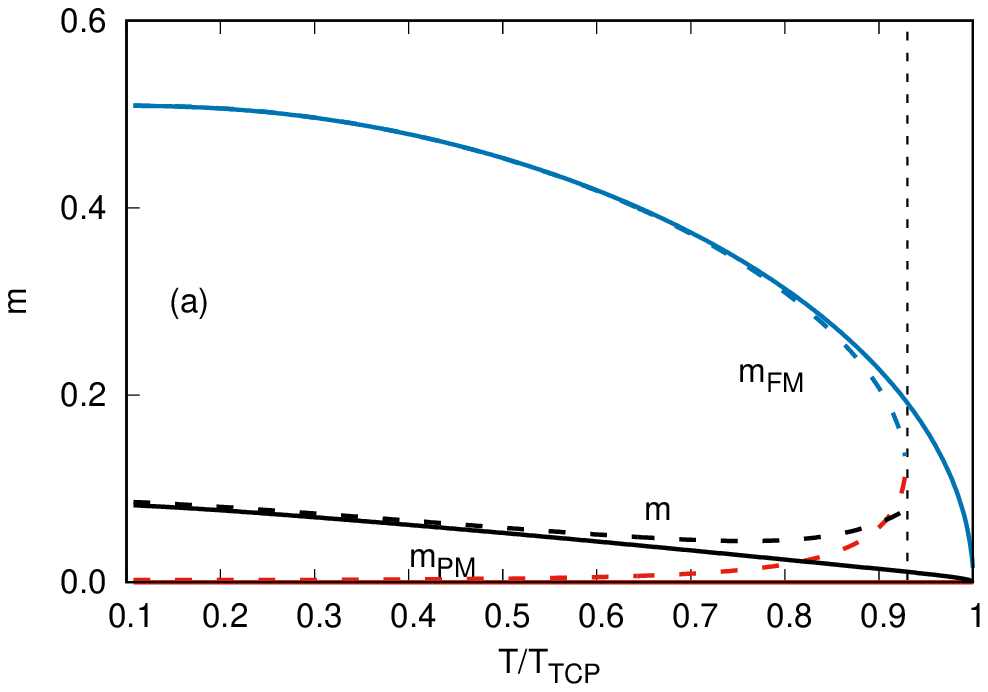}
 \includegraphics[angle=-90,clip=true,width=0.79\textwidth]{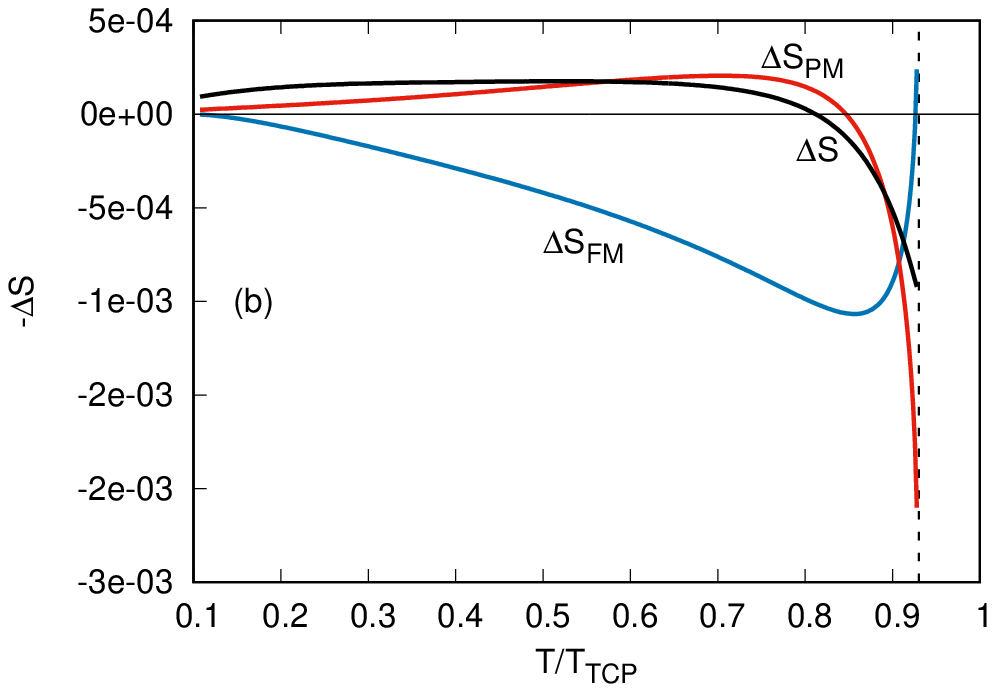}
\caption{\label{fig:T_vs_m,a_and_b}  
(a)~Temperature dependence of phase magnetization for ferromagnetic $m_{\rm FM}$ and paramagnetic $m_{\rm PM}$ phases involved in PS in zero magnetic field (solid lines) and finite magnetic field~(dashed lines). (b)~Temperature dependence of~phase entropy change $\Delta S_{\rm FM} $ and $\Delta S_{\rm PM}$. 
The temperature is taken in units of $T_{\rm TCP}$. 
The vertical dashed line shows the position $T^*(h)$. 
The sample average value of $\Delta S$ and $m$ for $n = n^\ast_0$ is shown for comparison. 
The parameters are the same as in Fig.~\ref{fig:T_vs_n}.
}
\end{figure}
A remarkable consequence of this is a perspective to~control the sign and magnitude of $\Delta S$ by changing the temperature in a narrow interval $ [T^{\ast}(h), T_{\rm TCP}] $, when $n\approx n^\ast_0$. 
When $n$ significantly differs from $n^\ast_0$ as the temperature increases, the system leaves PS area until a significant magnetic response is formed in the system.

Interestingly, the temperature dependencies  $\Delta S(T)$ and $m(T)$ for small $T$ are similar to that found for DyAl$_2$, see Ref.~\onlinecite{DyAl2}. Authors suppose that this anomaly is a consequence of an anisotropy and explains it within the Heisenberg model. 
The region of such dependence is limited by the temperature of the spin-flop transition. 
Here one can see an analogy with the results obtained, since the presence of two projections of magnetization along and perpendicular to the magnetic field in anisotropic systems is similar to the presence of two phases in phase-separated state of itinerant systems. 

\begin{figure}[ht]
    \includegraphics[angle=-90,clip=true,width=0.79\textwidth]{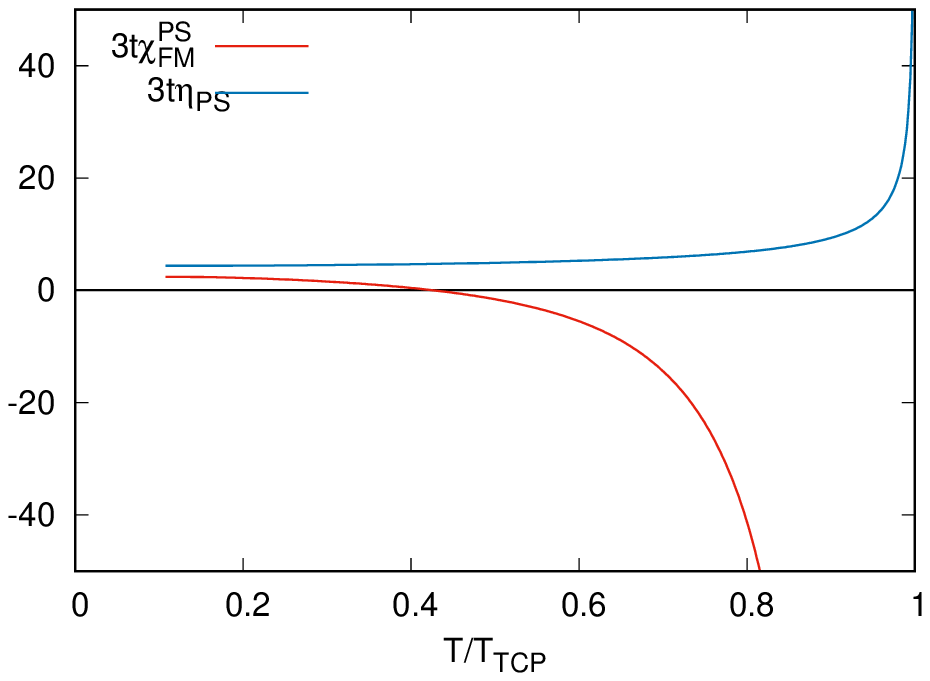}
\caption{
Temperature dependence of phase FM susceptibility $3t\cdot\chi^{\rm PS}_{\rm FM}$ and $3t\cdot\eta_{\rm PS}$ at FOPT line.  
The temperature is taken in units of $T_{\rm TCP}$. 
The parameters are the same as in Fig.~\ref{fig:T_vs_n}.
\label{fig:chi_eta}  
}
\end{figure}
Figure~\ref{fig:T_vs_m,a_and_b} shows the temperature dependence of $m$ and $\Delta S_\Phi$ calculated separately for~the~phases $\Phi$ involved in~PS in~the~case $U = 3t$. 
Note that within the PS region, these quantities by definition do not depend on $n$, whereas $n$ dependence of total $\Delta S$ originates from volume fraction $n$ dependence only, see~Eqs.~(\ref{eq:m_x}-\ref{eq:S_x}). 
For the case  $n = n^\ast_0$, temperature dependencies of total $m$ and $\Delta S$ (averaged over the sample volume) are also shown. 
For the ferromagnetic phase, the magnetic response (susceptibility) to the magnetic field is negative at $T > 0.43T_{\rm TCP}$~(see also~Fig.~\ref{fig:chi_eta} and derivation of this fact within the Landau theory in~Sec.~\ref{sec:Landau_theory}) and we have inverse MCE effect, $\Delta S_{\rm FM} > 0$, for almost all $T < T^*(h)$. 
For paramagnetic phase we have direct MCE effect $\Delta S_{\rm PM} < 0$ up to $T < 0.85T_{\rm TCP}$. 
For the case $T > 0.85T_{\rm TCP}$ we have rapid change of both phase $\Delta S_{\rm FM}$ and $\Delta S_{\rm PM}$ which is related with merging of $m_{\rm FM}$ and $m_{\rm PM}$ for phases involved in PS, see~Fig.~\eqref{fig:T_vs_m,a_and_b}. 

\begin{figure}[ht]
    \includegraphics[angle=-90,clip=true,width=0.79\textwidth]{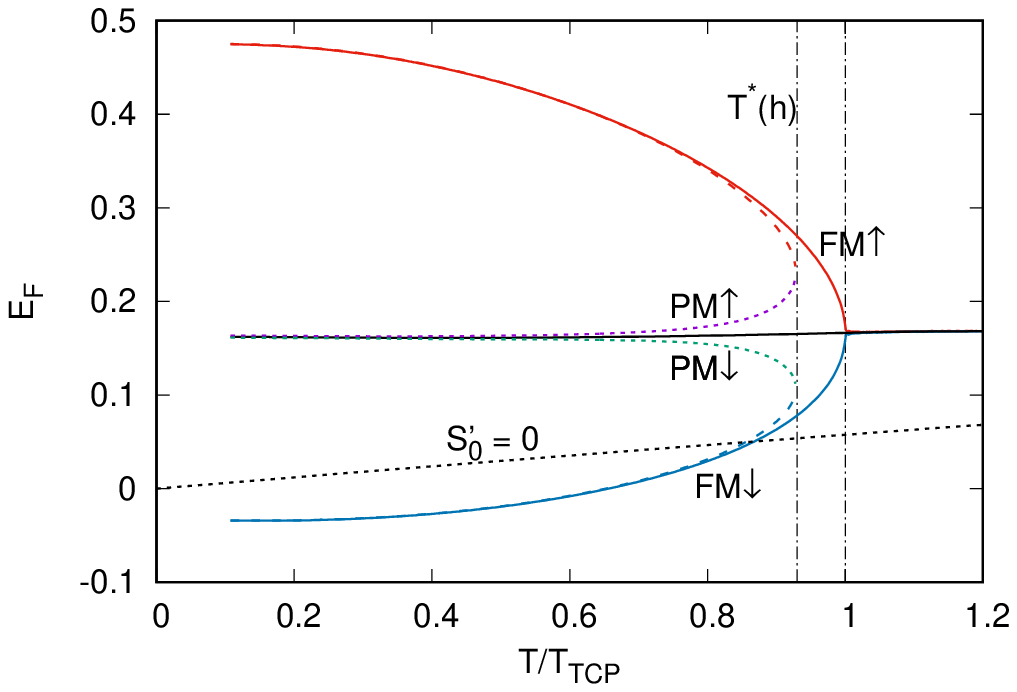}
\caption{
\label{fig:PS_curves_m_Ef}  
Subband effective dimensionless Fermi level $E^{\rm HFA}_{\rm F, \sigma}$ of FM and PM phase involved in PS in zero~(solid lines) and finite~(dashed lines)  magnetic field. 
Red (violet) lines correspond to FM (PM) phase $E^{\rm HFA}_{\rm F, \uparrow}$, blue (green) lines correspond to FM (PM) phase $E^{\rm HFA}_{\rm F, \downarrow}$.    
The parameters are the same as in~Fig.~\ref{fig:T_vs_n}. 
Black dotted line indicates the line $S_0'(T, E_{\rm F}) = 0$, see~Fig.~\ref{fig:Landau_PD}. 
Vertical dot-dashed line indicates the temperature bound of PS region in 
finite $T^*(h)$ magnetic field. 
The temperature is taken in units of $T_{\rm TCP}$. 
The parameters are the same as in Fig.~\ref{fig:T_vs_n}.
}
\end{figure}

Effective Fermi level of both spin subbands $E^{\rm HFA}_{\rm F,\sigma}$ at FOPT line, see Eq.~\eqref{eq:EF_sigma}, strongly depends on $T$.   Figure~\ref{fig:PS_curves_m_Ef} shows temperature dependence of $E^{\rm HFA}_{\rm F,\sigma}$ in zero and finite for both phases involved in PS.  
One can see that effective Fermi level for $\downarrow$-subband $E^{\rm HFA}_{\rm F,\downarrow}$ lies in the vicinity of van Hove singularity position $E = 0$. 

\begin{figure}[ht]
	\includegraphics[angle=-90,clip=true,width=0.69\textwidth]{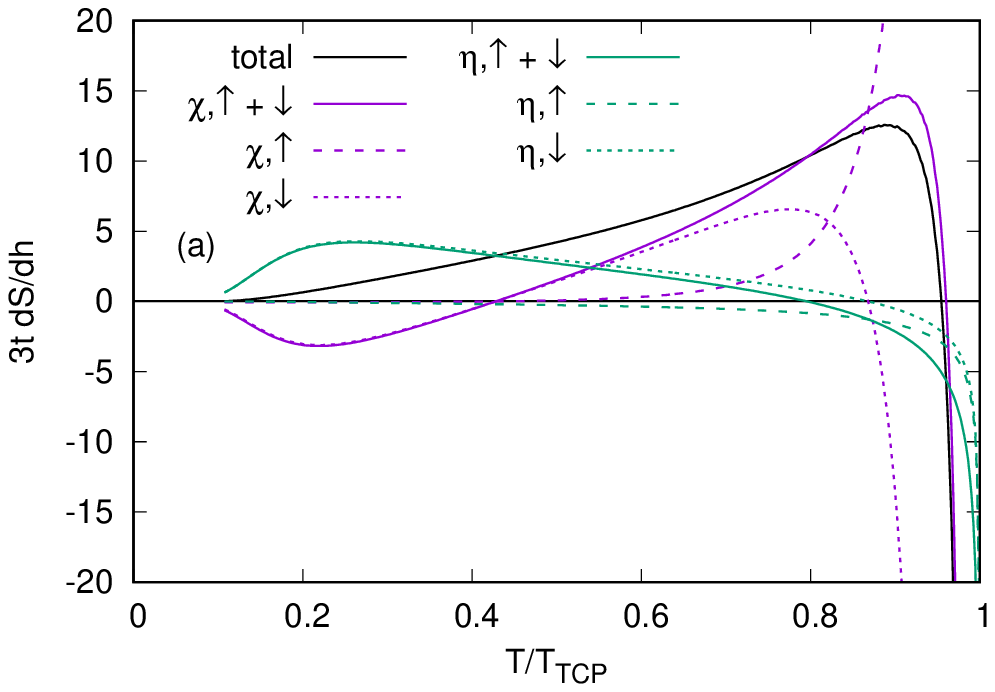}
    \includegraphics[angle=-90,clip=true,width=0.69\textwidth]{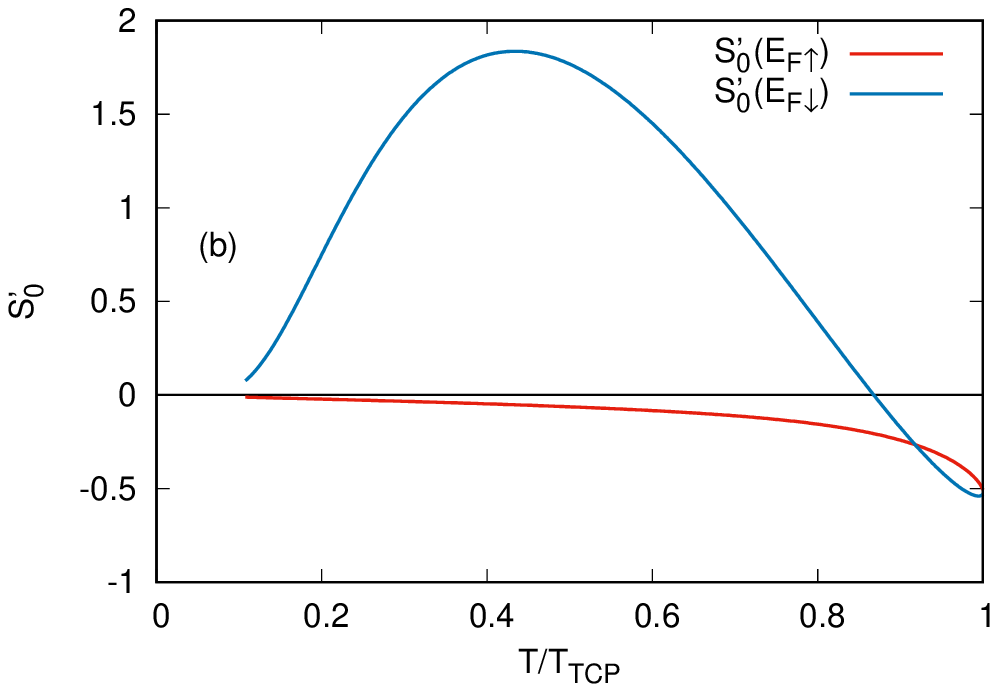}
\caption{
\label{fig:dS_dh}  
Temperature dependence of properties for~FM~phase involved in PS. 
(a)~The derivative $dS_{\rm FM}/dh$, see~Eq.~\eqref{eq:dS_dh_issue}~(black line) and separate contributions into this: violet (green) lines indicate first $\chi$~term (second $\eta$~term) in Eq.~\eqref{eq:dS_dh_issue}. Solid lines indicate sum of both spin projections contributions, dashed (dotted) line indicates spin up (down) projection. 
(b)~$S_0'(E^{\rm HFA}_{\rm F, \sigma})$ for both spin projections.  
The temperature is taken in units of $T_{\rm TCP}$. 
The parameters are the same as in Fig.~\ref{fig:T_vs_n}.
}
\end{figure}

A cause of inverse MCE for FM phase at FOPT line requires an  explanation. In general, the reason for this fact is that when the magnetic field is applied, the chemical potential changes, adjusting $n$ to into the phase separation region, cf.~discussion in~Sec.~\ref{sec:Landau_theory}. 
To reveal the origin of 
MCE sign dependence at FOPT line  
we calculate the magnetic field derivative of phase entropy for the phase $\Phi$
\begin{equation}\label{eq:dS_dh_issue}
\left(\frac{dS_\Phi}{dh}\right)_{\rm PS} = \frac{1}{\Pi^{\Phi}_{\uparrow}+\Pi^{\Phi}_{\downarrow} + 2U\Pi^{\Phi}_{\uparrow}\Pi^{\Phi}_{\downarrow}}\sum_\sigma\left(\gamma_\sigma(1+ U\Pi^{\Phi}_{\bar\sigma})\chi^{\rm PS}_\Phi
 + 2\eta_{\rm PS}\Pi^{\Phi}_{\bar\sigma}\right)S'_0(T,E^{\rm HFA}_{\rm F,\sigma}(T,n_\Phi,m_\Phi)),
\end{equation}
where $\eta_{\rm PS}$ is given by~Eq.~\eqref{eq:eta_1st_order} and
\begin{equation}\label{eq:dm_i/dh_PS}
\chi^{\rm PS}_\Phi  = \frac{\Pi^{\Phi}_{\uparrow} + \Pi^{\Phi}_{\downarrow} + 2U\Pi^{\Phi}_{\uparrow}\Pi^{\Phi}_{\downarrow}}{1-U^2\Pi^{\Phi}_{\uparrow}\Pi^{\Phi}_{\downarrow}} 
+ 
\frac{\Pi^{\Phi}_{\uparrow} - \Pi^{\Phi}_{\downarrow}}{1-U^2\Pi^{\Phi}_{\uparrow}\Pi^{\Phi}_{\downarrow}}\eta_{\rm PS},
\end{equation}
where $\Phi =$~FM, PM enumerates phases involved in PS and $\Pi^\Phi_{\sigma} = \Pi(T, E^{\rm HFA}_{\rm F,\sigma}(\mu, n_\Phi, m_\Phi))$. 
Equation~\eqref{eq:dS_dh_issue} contains two contributions: proportional to $\chi^{\rm PS}_\Phi$, $\chi$~term, and proportional to $\eta_{\rm PS}$, $\eta$~term, cf.~Eq.~\eqref{eq:dSdh_PS_state}.  
$T$~dependencies of $\left({dS_\Phi}/{dh}\right)_{\rm PS}$ and  $S'_0(T,E^{\rm HFA}_{\rm F,\sigma})$ for both spin projection in FM phase  involved in PS are shown in Fig.~\ref{fig:dS_dh}. 
Subband contributions are proportional to~$S_0'(T,E^{\rm HFA}_{\rm F,\sigma})$~(where a~prime here and below denotes $E_{\rm F}$ derivative); temperature dependence of the latter is shown in~Fig.~\ref{fig:dS_dh}(b). One can see that $S_0'$ dominates for $\sigma = \downarrow$ and positive. 
The strong dependence of $\left({dS_{\rm FM }}/{dh}\right)_{\rm PS}$ on~$T$ and the change in its sign are due to several facts: 
(i)~$|S'_0\left(T,E^{\rm HFA}_{\rm F,\uparrow}\right)| \ll S'_0\left(T,E^{\rm HFA}_{\rm F,\downarrow}\right)$,  (ii)~the~sign change~of~$S'_0(T,E^{\rm HFA}_{\rm F,\downarrow})$ at $T\approx 0.85T_{\rm TCP}$, (iii)~the~sign change of $\chi^{\rm FM}_{\rm PS}$ at $T \approx 0.43T_{\rm TCP}$, see~Fig.~\ref{fig:chi_eta}, cf.~the~discussion of  negative susceptibility of FM phase in~the~vicinity~of~tricritical point. 

At low temperature $\sigma = \downarrow$ contribution into $(dS_{\rm FM}/dh)_{\rm PS}$ dominates but $\chi$ and $\eta$~terms have opposite signs and almost cancel each other. $\downarrow$ $\chi$~term is negative since in this $T$ region both $S'_0(E_{\rm F,\downarrow}), \chi^{\rm PS}_{\rm FM} > 0$. 
At $T = 0.43T_{\rm TCP}$ $\chi^{\rm PS}_{\rm FM}$ and, in turn, $\chi$~term change their sign, and both $\chi$ and $\eta$~terms are positive which establishes inverse MCE for FM phase. 
$\eta_{\rm PS}$~everywhere holds its sign and the sign of $\eta$~term is fully determined by the sign of $S_0'(E_{\rm F,\downarrow})$.
Further $T$ increase results in dominating of $\chi$~term over $\eta$~term. 
In the vicinity of $T_{\rm TCP}$ $\downarrow$~term $\chi$ and $\eta$~term as well as $S_0'(E^{\rm HFA}_{\rm F,\downarrow})$ change their sign which results in sign change of $(dS_{\rm FM}/dh)_{\rm PS}$. The sign change of $S'_0(E_{\rm F})$ is associated with giant van Hove singularity of the density of states at~band bottom.

Now we see the reasons for the behavior of $\Delta S$ in Fig.~\ref{fig:T_vs_n}: the entropy of the paramagnetic phase decreases when the magnetic field is applied (excepting for $T$ in a close vicinity of~$T_{\rm TCP}$), while the entropy of the ferromagnetic phase increases, it is directly related with 
negative phase susceptibility of FM phase 
and impact of giant van Hove singularity of~the density~of~states.

For the considered case of giant van Hove singularity at the band bottom positioned in the vicinity of Fermi level,   
first-order phase transition FM-PM forces the electronic phase separation. The latter one leads to inverse MCE effect for the FM phase part of the sample. Since the FM part of the sample becomes more disordered, when magnetic field is applied, then its entropy increases. 
There is another one important point connected to the presence of giant van Hove singularity at the band bottom.
When temperature increases towards $T_{\rm TCP}$ the Fermi energy of the minority subband moves away from the giant van Hove singularity which results in change of the $\Delta S_{\rm FM}$ sign (see Figs.~\ref{fig:T_vs_m,a_and_b} and \ref{fig:PS_curves_m_Ef}).

\section{Conclusions}\label{sec:conclusions} 

In this work we study the first-order paramagnetic-ferromagnetic phase transition (FOPT) for an itinerant system with~fcc lattice and with giant van~Hove singularity in~the~electron density of states. The role of van Hove singularity is manyfold: (i) vHS is in general  necessary condition for ferromagnetic ordering in itinerant electron systems, (ii)~vHS results in first-order phase transition between paramagnetic and ferromagnetic phases with the corresponding electronic phase separation, (iii)~non-standard electronic properties caused by vHS provide, in turn, an~inverse entropy response on the application of magnetic field. It is shown that giant van Hove singularity, being a main cause of electronic phase separation, strongly affects temperature dependence of $\Delta S$ in the vicinity of~a~tricritical point. 

Chosen in this work perfect relation between nearest and next-nearest neighbor hopping integral values producing giant van Hove singularity for fcc-lattice is not strictly necessary. 
Earlier investigations of the problem yields that some deviations from this relation results in the replacement of divergence of DOS by a wide plateau with high value of $\rho(\epsilon)$, see Refs.~\onlinecite{2019:Igoshev_PMM,2019:Igoshev_JETP_letters,2023:Igoshev_fcc_fRG,2022:Igoshev_DOS_fcc}. 
This fact does not break completely anomalous thermodynamical properties which allows one to expect the PS existence and inverse MCE in this case. 

We consistently treat appearing near the FOPT electronic phase separation and its~impact on a value of~magnetocaloric effect. The thermodynamic theory based on Landau grand potential expansion  for the ferromagnetic-paramagnetic phase transition is developed for the electronic phase-separated state. In contrast to conventional free energy expansion the proposed grand potential expansion provides correct account for electronic phase separation effects.
It is rigorously shown that ferromagnetic phase involved in the phase-separated state exhibits negative magnetic susceptibility in the vicinity of~tricritical point. It leads to the fact that FM part of the sample becomes more disordered, when magnetic field is applied, then its entropy increases, so an inverse MCE effect manifests itself.

The electronic phase separation and magnetocaloric effect (MCE) are considered within the Hubbard model for~face-centered cubic (fcc) lattice with giant van Hove singularity of electron density of states at the band bottom. 
Within the Hartree-Fock approximation it is shown that such model of itinerant magnet exhibits the~first-order ferromagnet-paramagnet phase transition~(FOPT)  with electronic phase separation and inverse magnetocaloric effect deep inside the phase-separated region. 
The Hartree-Fock approximation used can be in principle replaced by any better approximation. The formulated in the paper general statements are valid for~any itinerant system undergoing a first-order magnetic phase transition, which always leads to an unusual behavior of $\Delta S$ for the case when the electron filling lies in the PS filling interval. 
However electronic phase separation characteristics depend on approximation applied which leads to quantitative differences only. 

Note that the long-range Coulomb energy not taken into account within the approximation used, cannot qualitatively change the results: as scanning tunneling microscopy experiments have shown, see e.~g.~Ref.~\onlinecite{STM}, in such cases the system undergoes nanoscale phase separation (a specific implementation of the heterogeneity configuration is set up by the location of impurities and defects), which corresponds to a balance between inner system energy taken into account within our investigation and Coulomb long-range interaction. 

Temperature dependence of $\Delta S$ for the mean-field solution of the non-degenerate Hubbard model is analyzed in detail for different band filling values. 
In general, it is shown that the MCE in itinerant systems undergoing a PM-FM FOPT can have remarkable characteristics due to presence of phase separation---strong band filling dependence of $\Delta S$ and kinks in its temperature dependence. In particular, it was found that the contribution to $\Delta S$ from the ferromagnetic phase in the low-temperature region is positive. 
With increase of temperature, this contribution has pronounced peak
at the temperature of the PS region leaving for a finite magnetic field and has a strong almost linear dependence inside the PS region at zero magnetic field.
To get this effect, it is sufficient to provide special behavior of the density of states only (e.g. the presence of van Hove singularities near the Fermi level) and no additional interactions are required. 
The possibility to control $\Delta S$ sign by changing both temperature and band filling of magnetocaloric materials is important to interpret a lot of experimental data, possible technological applications and further theoretical developments.

The description of MCE within a proposed generalized Landau theory can acquire perspective extensions taking into account additional interactions such as magnetoelastic coupling~\cite{2006:Podgornyh}, spin fluctuations~\cite{2007:Igoshev} etc.  
For example, since magnetoelastic interaction couples electronic and lattice subsystems, structural FOPT transition can trigger electronic phase separation or vice versa.

\section{Acknowledgments}
We are grateful to V.~Yu.~Irkhin, A.~N.~Ignatenko, N.~S.~Pavlov for fruitful discussions. 
The theoretical studies in Section~2 is supported by the state assignment of the Ministry of Science and Higher Education of the Russian Federation (theme ``Quant'' No. 122021000038-7) for development of generalized Landau functional approach. 
The theoretical studies in Section 3 are in part supported by the Russian Science Foundation (Project No. 24-43-00156) for entropy calculations within Hartree-Fock approximation for the Hubbard model. 


\end{document}